\documentclass[prc,twocolumn,floatfix,groupedaddress,nofootinbib,showpacs,preprintnumbers,
amsmath,amssymb,amsfonts,superscriptaddress,widetable] {revtex4-1}
\usepackage{bm}
\usepackage{mathrsfs}
\usepackage{amssymb}
\usepackage{amsmath}
\usepackage{graphicx}
\usepackage{array}
\usepackage[normalem]{ulem}
\usepackage{xcolor}
\newcommand{\D}{{\mathrm d}}

\begin{document}
\title{Quantum Nuclear Pasta and Nuclear Symmetry Energy}
\author{F. J. Fattoyev}\email{ffattoye@indiana.edu}
\affiliation{Center for Exploration of Energy and Matter and
                  Department of Physics, Indiana University,
                  Bloomington, IN 47405, USA}
\author{C. J. Horowitz}\email{horowit@indiana.edu}
\affiliation{Center for Exploration of Energy and Matter and
                  Department of Physics, Indiana University,
                  Bloomington, IN 47405, USA}
\author{B. Schuetrumpf}\email{schutrum@nscl.msu.edu}
\affiliation{FRIB Laboratory, Michigan State University, East
                Lansing, Michigan 48824, USA}

\date{\today}
\begin{abstract}
Complex and exotic nuclear geometries are expected to appear
naturally in dense nuclear matter found in the crust of neutron
stars and supernovae environment collectively referred to as
``nuclear pasta". The pasta geometries depend on the average baryon
density, proton fraction and temperature and are critically
important in the determination of many transport properties of
matter in supernovae and the crust of neutron stars. Using a set of
self-consistent microscopic nuclear energy density functionals we
present the first results of large scale quantum simulations of
pasta phases at baryon densities $0.03 \leq \rho \leq 0.10$
fm$^{-3}$, proton fractions $0.05 \leq Y_{\rm p} \leq 0.40$, and
zero temperature. The full quantum simulations, in particular, allow
us to thoroughly investigate the role and impact of the nuclear
symmetry energy on pasta configurations. We use the \verb"Sky3D"
code that solves the Skyrme Hartree-Fock equations on a
three-dimensional Cartesian grid. For the nuclear interaction we use
the state of the art \mbox{\tt UNEDF1} parametrization, which was
introduced to study largely deformed nuclei, hence is suitable for
studies of the nuclear pasta. Density dependence of the nuclear
symmetry energy is simulated by tuning two purely isovector
observables that are insensitive to the current available
experimental data. We find that a minimum total number of nucleons
$A=2000$ is necessary to prevent the results from containing
spurious shell effects and to minimize finite size effects. We find
that a variety of nuclear pasta geometries are present in the
neutron star crust and the result strongly depends on the nuclear
symmetry energy. The impact of the nuclear symmetry energy is less
pronounced as the proton fractions increase. Quantum nuclear pasta
calculations at $T=0$ MeV are shown to get easily trapped in
meta-stable states, and possible remedies to avoid meta-stable
solutions are discussed.
\end{abstract}

\smallskip
 \maketitle

\section{Introduction}
\label{Introduction}

The baryon matter in the Universe organizes itself based on the
short range nuclear attraction and the long-range Coulomb repulsion.
At densities much lower than the nuclear saturation density, $\rho_0
\approx 0.16$ fm$^{-3}$, the nuclear and atomic length scales are
well separated, so nucleons bind into nuclei that, in turn, are
segregated in a Coulomb lattice. All terrestrial materials as well
as the matter in the outer layers of the neutron star crust are
expected to harbor such sites. However, the density of matter inside
the neutron star crust---as well as in the regions of
supernov\ae\---has a range that spans several orders of magnitude.
In high-density regions, $\rho \gtrsim \rho_0$, which are expected
in the core of neutron stars, the short range nuclear interaction
significantly dominates over the atomic length scales and the matter
assumes a uniform phase. At sub-saturation baryon densities, $0.1
\rho_0 \lesssim \rho \lesssim 0.8 \rho_0$, a region expected at the
bottom layers of the inner crust, these two length scales become
comparable, and the matter develops complex and exotic structures as
a result of the so-called \emph{Coulomb frustration}. In this case,
there is a strong competition between the Coulomb and the strong
interaction, which leads to the emergence of various complex
structures with similar energies that are collectively referred to
as ``nuclear pasta". A significant progress has been made in
simulating this exotic region\,\cite{Watanabe:2003xu,
Watanabe:2004tr, Horowitz:2004yf, Horowitz:2004pv, Horowitz:2005zb,
Schneider:2013dwa}, since their initial prediction over several
decades ago\,\cite{Ravenhall:1983uh,Hashimoto:1984,Lorenz:1992zz}.
There are ongoing efforts aiming to determine the possible shapes of
the nuclear pasta\,\cite{Horowitz:2014xca}, as it is believed that
the elastic and transport properties---such as electrical and
thermal conductivities, shear and bulk viscosities---of nuclear
pasta play crucial role for thermal evolution, magnetic field
evolution, rotation, and oscillations of neutron
stars~\cite{Lorenz:1992zz, Levin:2000vq, Gusakov:2004mj,
Gearheart:2011qt, Pons:2013nea}. Moreover, they can significantly
impact neutrino opacities in the core-collapse supernov\ae, which in
turn strongly influences the dynamics of the core collapse and the
cooling of proto-neutron
stars~\cite{Bethe:1990mw,Horowitz:2004yf,Alloy:2010fk,Janka:2012wk}.
In this paper we will investigate large scale quantum simulations of
nuclear pasta phases at baryon densities $0.03 \leq \rho \leq 0.10$
fm$^{-3}$, proton fractions $0.05 \leq Y_{\rm p} \leq 0.40$, and
zero temperature by using a set of self-consistent microscopic
nuclear energy density functionals, and discuss the role and impact
of the nuclear symmetry energy.

The traditional approach to study nuclear pasta phases often
involves symmetry arguments to determine what is the most favored
structure at a given baryon density, $\rho$, temperature, $T$, and
proton fraction, $Y_{\rm p}$. The system is then minimized by either
adding an external guiding potential or with some other sorts of
biased initialization that explicitly makes assumptions about the
geometrical shapes of the nuclear pasta. Some example model
calculations include the use of the liquid-drop
model~\cite{Ravenhall:1983uh, Pethick:1998qv, Nakazato:2009ed},
Thomas-Fermi and Wigner-Seitz cell
approximations~\cite{Lorenz:1992zz, Williams:1985prf, Lassaut:1987,
Oyamatsu:1993zz, Maruyama:2005vb}. Perhaps some of the most exotic
phases obtained using pre-assumed shapes are the gyroid and diamond
morphologies\,\cite{Nakazato:2009ed, Schuetrumpf:2014aea}. There are
other approaches that do not explicitly assume any shape for the
nuclear pasta phase. These include calculations based on the
Thomas-Fermi approximation\,\cite{Williams:1985prf, Okamoto:2011tc,
Magierski:2001ud, Grill:2012tp}, non-relativistic Skyrme
Hartree-Fock methods\,\cite{Newton:2009zz, Schuetrumpf:2012cj,
Pais:2012js, Sagert:2015rra}, relativistic density-functional
theory\,\cite{Maruyama:2005vb}, relativistic mean
field-approximation\,\cite{Gupta:2013rda, Avancini:2008zz,
Avancini:2008kg}, quantum molecular dynamics
(QMD)\,\cite{Maruyama:1997rp, Watanabe:2002sf, Watanabe:2003xu,
Watanabe:2004tr, Watanabe:2005qt, Sonoda:2007sx, Watanabe:2009vi}
and semi-classical molecular dynamics (MD)\,\cite{Horowitz:2004yf,
Horowitz:2004pv, Horowitz:2005zb, Horowitz:2008vf,
Piekarewicz:2011qc, Dorso:2012pc, Caplan:2014gaa, Horowitz:2014xca,
Schneider:2014lia, Horowitz:2015gda, Schneider:2016zyx} simulations.
Recently using MD simulations more exotic structures have also been
identified, such as flat plates with a lattice of holes, termed as
``nuclear waffles"\,\cite{Schneider:2014lia}, and flat plates that
are connected by spiral ramps\,\cite{Horowitz:2015gda}.

For small systems, these studies are often performed in the unit
cell filled with neutrons, protons and electrons alongside the
specific symmetry assumptions and boundary conditions. The pasta
matter is then described as a lattice made of a large number of
these unit cells. When performing numerical studies, it is important
to consider the non-trivial role of the simulation volume. Since
only periodic geometries that fit into the unit cell can be
explored, the simulation space must be sufficiently large to contain
at least one unit cell of the pasta structure. Even if this
condition is fulfilled, finite size effects such as dependence on
the geometry of the simulation
space\,\cite{GimenezMolinelli:2014dha} and numerical shell
effects\,\cite{Newton:2009zz} may appear. As a result, the
simulation volume needs to be maximized to ensure that finite size
effects are minimal. Advances in computational power in the last
decade have allowed for sophisticated fully self-consistent
calculations by using Skyrme Hartree-Fock (SHF) calculations at
finite temperature\,\cite{Newton:2009zz, Schuetrumpf:2012cj,
Pais:2012js, Sagert:2015rra}. Whereas these computations showed a
richer variety of pasta shapes than the original five
geometries\,\cite{Hashimoto:1984}, consistent with results obtained
by the MD simulations that use significantly larger simulation
volumes, they are typically reproduced by assuming various symmetry
arguments or an \textit{a priori} assumed final pasta shapes.
Moreover, due to the limitation of computational power, these
calculations were often limited to a single periodic structure,
therefore leading to the pasta shapes that may exhibit significant
dependence on the finite size of the simulation box. Therefore, it
is necessary to perform quantum simulations with a much larger
number of nucleons to overcome finite size effects, as well as to
minimize various numerical effects coming from different symmetry
considerations. The progress in the high-performance computing in
recent years allows us to take further steps in this direction,
which is the main topic of this manuscript. Indeed, the recent
decadal nuclear survey\,\cite{NuclearPhysics:2013} puts forward that
``high performance computing provides answers to questions that
neither experiment nor analytic theory can address; hence, it
becomes a third leg supporting the field of nuclear physics.''

Calculations with more than a few thousand nucleons so far were only
manageable by considerably simplifying the nuclear interaction. That
is what was done in previous works that study nuclear pasta using
classical or quantum MD simulations. The advantage of MD simulations
lies in their ability to simulate large systems where the length of
the simulation space is several hundred fermis, and therefore
significantly exceeds the size of a unit cell. This allows to study
pasta structures that are less bound to the geometry and boundary
conditions of the simulation volume. However, although MD approaches
can include quantum effects qualitatively, the nuclear interaction
is typically given by a schematic two-body potential. For
self-consistent quantum calculations that account for Pauli
blocking, spin-orbit forces and nucleon pairing, simulations using
microscopic energy density density functionals (EDF) in the form of
SHF are usually performed. As mentioned above the current drawback
of these methods is their high computational cost. As a consequence,
the size of the system is typically chosen to be much smaller than
the one for MD methods. By using nuclear configurations that
conserve reflection symmetry in the three Cartesian directions,
Ref.\,\cite{Newton:2009zz} were able to simulate effectively larger
quantum systems by performing the computation only in one octant of
the unit cell. In this study we will not restrict our simulation
with nuclear configurations that assume any kind of spatial
symmetries. In particular, we will perform quantum simulations of
nuclear pasta using Skyrme Hartree-Fock model\,\cite{Maruhn:2013mpa}
with no pre-assumed pasta geometries, and we will address the
following main questions:
\begin{itemize}
\item[(a)] what is the minimum size of the simulation volume necessary to minimize finite size effects?
\item[(b)] what is the role of nuclear symmetry energy in the nuclear pasta
formation in neutron star crust and supernovae?
\item[(c)] how does the initial configuration of the system impact on the
converged pasta structure?
\end{itemize}

We have organized the paper as follows. In Sec.\,\ref{Formalism} we
review the essential details required to simulate nuclear pasta.
First, we modify the density dependence of the symmetry energy in
the Skyrme force interaction \mbox{\tt UNEDF1} by adapting two
purely isovector parameters. We present predictions for the ground
state properties of several closed-shell finite nuclei using the
original and the modified parametrizations. Then, we discuss the
impact of the grid spacing, accuracy considerations, and optimum
simulation runtime. Special attention is paid to the impact of
finite size effects, in which we identify the minimum simulation
volume that contains at least one period of the pasta structure. In
Sec.\,\ref{Results} we discuss the outcomes of our results. First,
we provide predictions for nuclear pasta with low proton fractions
corresponding to the crust of neutron stars. Second, we explore a
range of proton fractions corresponding to the matter found in
supernovae. Last, we discuss the non-trivial effect of initial
configurations on the final pasta configuration. Finally, we offer
our conclusions in Sec.\,\ref{Conclusions}.

\section{Formalism}
\label{Formalism}

\subsection{Nuclear Interaction and Symmetry Energy}
\label{NucIntNSE}

To simulate the nuclear pasta structures we use the publicly
available Skyrme TDHF code \verb"Sky3D" that solves the static
Skyrme Hartree-Fock equations in a three-dimensional Cartesian mesh
with a damped gradient iteration method on an equi-distant grid and
without symmetry restrictions\,\cite{Maruhn:2013mpa}. For the nuclear
pasta simulations we use periodic boundary conditions that also
includes a homogeneous negative electron background to ensure the
charge neutrality of the system. This so-called \textit{jellium
approximation} is suitable for the nuclear pasta studies as they are
expected to be present in charge neutral environments, such as in
the inner crust of neutron stars. A screened Coulomb interaction is
not considered, as its influence should be very small for the box
lengths considered in our study\,\cite{Alcain:2013bza}. For a full
description of the code and the Skyrme Hartree-Fock method we refer
the reader to Ref.\,\cite{Maruhn:2013mpa}.

For the nuclear interaction we select a state-of-the-art Energy
Density Functional (EDF) of Skyrme type. The total energy is given
by
\begin{equation}
E_{\rm tot} = E_{\rm kin} + E_{\rm Sk} + E_{\rm C} \ ,
\end{equation}
where $E_{\rm kin}$ is the kinetic energy, $E_{\rm Sk}$ is the
Skyrme contribution, and $E_{\rm C}$ is the Coulomb contribution.
The Skyrme energy function contains five contributions
\begin{equation}
E_{\rm Sk} = E_0 + E_1 + E_2 + E_3 + E_{\rm ls} \ ,
\end{equation}
which are in detail
\begin{subequations}
    \begin{eqnarray}
      \label{eq:ehf0}
      E_0&=&\int \D^3\!r\,\left(\frac{b_0}{2}\rho^2-\frac{b_0'}{2}\sum_q\rho_q^2\right) \ , \\
      \label{eq:ehf1}
      E_1&=&\int \D^3\!r\,\left(b_1\rho\tau
        -b_1'\sum_q\rho_q\tau_q\right) \ , \\
      \label{eq:ehf2}
      E_2&=&\int \D^3\!r\,\left(-\frac{b_2}{2}\rho\Delta\rho
        +\frac{b_2'}{2}\sum_q\rho_q\Delta\rho_q\right) \ , \\
      \label{eq:ehf3}
      E_{3}&=&\int \D^3\!r\,\left(\frac{b_3}{3}\rho^{\alpha+2}
        -\frac{b_3'}{3}\rho^\alpha\sum_q\rho_q^2\right) \ , \\
     \label{eq:ehfls}
      E_{\rm ls}&=&\int \D^3\!r\,\left(-b_4\rho \nabla \cdot \vec{J}
        -b_4'\sum_q\rho_q \nabla \cdot \vec{J}_q \right)
    \end{eqnarray}
\end{subequations}
for time-independent calculations, where $E_0$ is known as the
zero-range term, $E_1$ as the effective mass term, $E_2$ as the
finite-range term, $E_3$ as the density dependent term, and $E_{\rm
ls}$ as the spin-orbit term. Here $\rho$ is the total particle
density, $\tau$ is the total kinetic density and $\vec{J}$ is the
total spin-orbit density, and if a subscript $q$ is present it
labels the densities of either neutrons or protons. The Coulomb
energy $E_{\rm C}$ consists of the standard expression for a charge
distribution in its own field plus the exchange term in the Slater
approximation:
\begin{equation}
     \label{eq:eC}
E_{\rm C} = \frac{e^2}{2} \int \D^3\!r \D^3\!r^{\prime}
\frac{\rho_{p}(\vec{r}) \rho_{p}(\vec{r}^{\prime})
}{\vec{r}-\vec{r}^{\prime}} -
\frac{3^{\frac{4}{3}}e^2}{4\pi^{\frac{1}{3}}} \int \D^3\!r
\rho_p^{\frac{4}{3}}(\vec{r}) \ ,
\end{equation}
where $e$ is the elementary charge. The parameters $\alpha$, $b_i$
and $b_i'$, $i\in \{0,1,2,3\}$ are fitted to experimental data. We
chose the \mbox{\tt UNEDF1} parametrization, whose parameters were
fitted to a selected set of nuclear masses, charge radii, odd-even
mass differences, and the experimental excitation energies of
fission isomers in the actinides\,\cite{Kortelainen:2011ft}. Given
that \mbox{\tt UNEDF1} was introduced to better study largely
deformed nuclei, we find this Skyrme force very suitable for our
studies of nuclear pasta that can take shapes of strongly elongated
nuclei, in particular.

Although the current extensive experimental database is sufficient
to constrain most of the parameters of the nuclear interaction, many
nuclear forces widely disagree in their description of the isovector
channel of the nuclear force due to poorly constrained isovector
parameters. In the realm of nuclear matter this means that the
density dependence of the nuclear symmetry energy remains poorly
determined. Since nuclear pasta is expected to form in a
neutron-rich environment, the role of the nuclear symmetry energy on
the pasta formation and the pasta phase transition needs to be
thoroughly analyzed. It has been shown by Oyamatsu and Iida that
pasta formation may not be universal in the neutron star crust and
that its existence is intimately related to the density dependence
of the symmetry energy\,\cite{Oyamatsu:2006vd}, where the pasta
regime was predicted to appear when the density slope of the
symmetry energy is $L \lesssim 100$ MeV (see
Ref.\,\cite{Horowitz:2014bja} for definitions of symmetry energy
parameters). Recently there have been several studies in the context
of the Thomas-Fermi approximation that analyzed the impact of
density slope of the nuclear symmetry energy $L$ on the pasta phase
structure\,\cite{Grill:2012tp, Bao:2014sma, Bao:2015cfa}. In
particular, it was found that whereas models with small value of $L$
exhibit a variety of pasta structures, most of these structures are
faded away when one considers models with the large value of $L$
corresponding to the stiff nuclear symmetry energy.

Intensive efforts have been devoted to constrain the density
dependence of the nuclear symmetry energy in recent years from using
various approaches (please see Refs.\,\cite{Tsang:2012se,
Lattimer:2012nd, Li:2013ola} and references therein). These efforts
have recently led to a close convergence of the value of symmetry
energy at saturation being around $J \approx 30$ MeV and its density
slope of $L \approx 60$ MeV. Nevertheless, the associated error-bars
from different approaches vary broadly and the possibility that $J$
and $L$ parameters can be significantly different from these
currently inferred values cannot be ruled out\,\cite{Li:2013ola}.
For this reason, we have modified two purely isovector parameters of
the \mbox{\tt UNEDF1} by following the tuning scheme as described in
Ref.\,\cite{Fattoyev:2012ch}. In particular, we modify the Skyrme
parameters $x_0$ and $x_3$ (Table \ref{Table1}), that in turn modify
the parameters $b_0$, $b_0'$, $b_3$ and $b_3'$ of the EDF (Eqs.
(\ref{eq:ehf0}) and (\ref{eq:ehf3})) which are given by
\begin{equation}
  \label{eq:force-coeff}
  \begin{split}
    b_0&=t_0\,\left(1+\tfrac1{2}x_0\right) \\
    b_0'&=t_0\,\left(\tfrac1{2}+x_0\right)\\
    b_3&=\tfrac1{4}t_3 \left(1+\tfrac1{2}x_3\right) \\
    b_3'&=\tfrac1{4}t_3 \left(\tfrac1{2}+x_3\right) \\
  \end{split}
\end{equation}
in terms of the Skyrme parameters $t_0$, $t_3$, $x_0$ and $x_3$. The
tuning method allows one to generate a family of model interactions
that are almost indistinguishable in their predictions for a large
set of the nuclear ground state observables that are mostly
isoscalar in nature, yet predict different isovector observables. As
a contrast to the original \mbox{\tt UNEDF1} model that has a
relatively soft symmetry energy with $L = 40$ MeV, we generated a
model that predicts a rather stiff symmetry energy of $L=80$ MeV.

\begin{table*}[t]
\begin{tabular}{|l||c|c|c|c|c|c|c|c|r|}
 \hline
 Model                        &$x_0$             & $x_3$  & $\rho_{0}$ & $\varepsilon_0$ & $K_0$ & $Q_0$ & $J$   & $L$   & $K_{\rm sym}$ \\
 \hline
 \hline
 \mbox{\tt UNEDF1}                &  $+$0.0537569200 & $-$0.1624911700 & 0.1587 & -15.80 & 220.0 & -405.0 & 28.99 & 40.00 & $-$179.5 \\
 \mbox{\tt UNEDF1}$^{\star}$      &  $-$0.3237259090 & $-$0.7725758299 & 0.1587 & -15.80 & 220.0 & -405.0 & 32.87 & 80.00 & $-$71.42  \\

\hline
\end{tabular}
\caption{Bulk parameters characterizing the behavior of neutron-rich
matter around saturation density $\rho_{0}$. Here $x_0$ and $x_3$
are the two pure isovector parameters of the Skyrme force \mbox{\tt
UNEDF1} that have been re-fitted to obtain an interaction with the
stiff symmetry energy, \mbox{\tt UNEDF1}$^{\star}$. The quantities
$\varepsilon_0$, $K_0$, and $Q_0$ represent the binding energy per
nucleon, incompressibility coefficient, and the ``skewness"
coefficient of symmetric nuclear matter at $\rho_{0}$ . Similarly,
$J$, $L$, and $K_{\rm sym}$ represent the energy, slope, and
curvature of the symmetry energy at saturation density. All
quantities are in MeV, except for $\rho_{0}$ which is given in
fm$^{-3}$. A detailed explanation of all these quantities may be
found in Ref.\,\cite{Piekarewicz:2008nh}.} \label{Table1}
\end{table*}

  \begin{table}[t]
  \begin{tabular}{|c|c|c|c|c|c|}
    \hline
    Nucleus & Observable & Experiment & $L=40$ MeV & $L=80$ MeV  \\
    \hline
    \hline
    ${}^{16}$O & $B/A$~(MeV)   & $-7.98$ & $-7.56$ & $-7.56$   \\
     & $r_{\rm ch}$~(fm)       & $\phantom{-}2.70$ & $\phantom{-}2.81$ & $\phantom{-}2.81$   \\
     & $r_{\rm skin}$~(fm)     &  ---  & $-0.02$ & $-0.02$   \\
   \hline
    ${}^{40}$Ca & $B/A$~(MeV)  & $-8.55$ & $-8.52$ & $-8.52$   \\
     & $r_{\rm ch}$~(fm)       & $\phantom{-}3.48$ & $\phantom{-}3.50$ & $\phantom{-}3.50$   \\
     & $r_{\rm skin}$~(fm)     &  ---  & $-0.04$ & $-0.04$   \\
   \hline
    ${}^{48}$Ca & $B/A$~(MeV)  & $-8.67$ & $-8.60$ & $-8.61$   \\
     & $r_{\rm ch}$~(fm)       & $\phantom{-}3.47$ & $\phantom{-}3.53$ & $\phantom{-}3.52$   \\
     & $r_{\rm skin}$~(fm)     &  ---  & $\phantom{-}0.18$ & $\phantom{-}0.21$   \\
   \hline
    ${}^{90}$Zr & $B/A$~(MeV)  & $-8.71$ & $-8.72$ & $-8.72$   \\
     & $r_{\rm ch}$~(fm)       & $\phantom{-}4.27$ & $\phantom{-}4.28$ & $\phantom{-}4.28$   \\
     & $r_{\rm skin}$~(fm)     &  ---  & $\phantom{-}0.08$ & $\phantom{-}0.10$   \\
   \hline
    ${}^{132}$Sn & $B/A$~(MeV) & $-8.35$ & $-8.35$ & $-8.33$   \\
     & $r_{\rm ch}$~(fm)       &  ---  & $\phantom{-}4.72$ & $\phantom{-}4.72$   \\
     & $r_{\rm skin}$~(fm)     &  ---  & $\phantom{-}0.25$ & $\phantom{-}0.30$   \\
   \hline
    ${}^{208}$Pb & $B/A$~(MeV) & $-7.87$ & $-7.88$ & $-7.86$   \\
     & $r_{\rm ch}$~(fm)       & $\phantom{-}5.50$ & $\phantom{-}5.51$ & $\phantom{-}5.51$   \\
     & $r_{\rm skin}$~(fm)     &  ---  & $\phantom{-}0.18$ & $\phantom{-}0.23$   \\
   \hline
  \end{tabular}
 \caption{Experimental data (where available) and theoretical predictions for the two EDFs
 for the binding energy per nucleon, charge radii and neutron skin thickness for several closed shell nuclei.}
 \label{Table2}
 \end{table}
In Table\,\ref{Table1} we present the nuclear matter bulk parameters
for these two interactions. And in Table\,\ref{Table2} we show the
success of such tuning by presenting predictions for binding
energies and charge radii of several closed shell nuclei. We also
present the corresponding neutron skin thicknesses $r_{\rm skin}$ of
these nuclei. It is worth mentioning that the measurement of $r_{\rm
skin}$ in $^{48}$Ca and $^{208}$Pb are of enormous significance due
to their very strong correlation to the slope of the symmetry energy
around saturation density\,\cite{Brown:2000pd, Furnstahl:2001un,
Centelles:2008vu, RocaMaza:2011pm}. The neutron skin thickness of
$^{208}$Pb has been preliminarily measured by the PREX Collaboration
at Jefferson Laboratory\,\cite{Abrahamyan:2012gp}, and will be
measured with higher accuracy by the PREX-II
experiment\,\cite{PREXII} in 2017. An already approved CREX
experiment on the other hand aims to measure the neutron skin
thickness of $^{48}$Ca\,\cite{CREX}. The calculations of Table II is
performed using the \verb"Sky3D" code with isolated boundary
conditions, for the Coulomb force. The charge radius is calculated
using the point-proton mean-square radius $\langle r^2 \rangle_{\rm
pp}$ from \verb"Sky3D" and the approximate analytic
formula\,\cite{Ong:2010gf}
\begin{equation}
\label{ChargeRMS} \langle r^2 \rangle_{\rm ch} = \langle r^2
\rangle_{\rm pp} + \langle R_{\rm p}^2 \rangle + \frac{N}{Z} \langle
R_{\rm n}^2 \rangle + \frac{3}{4M^2} + \langle r^2 \rangle_{\rm so}
\ ,
\end{equation}
where $\langle R_{\rm p}^2 \rangle= 0.7658$ fm$^2$ and $\langle
R_{\rm n}^2 \rangle = -0.1161$ fm$^2$ are the mean-square charge
radii of the proton and the neutron, respectively, $\frac{3}{4M^2} =
0.03312$ fm$^2$ is the so-called Darwin-Foldy term, and $\langle r^2
\rangle_{\rm so}$ is the relativistic spin-orbit correction. Notice,
that the slope of the symmetry energy  $L$ is closely related to the
pressure of pure neutron matter at saturation density,
\textit{i.e.}, $L \approx 3 P(\rho_0)/\rho_0$. Therefore the larger
$L$ results in the higher neutron pressure which leads to greater
neutron radii and thicker neutron skins as neutrons are pushed out
against surface tension.

\begin{figure}[tbh]
\vspace{-0.05in}
\includegraphics[width=1.0\columnwidth,angle=0]{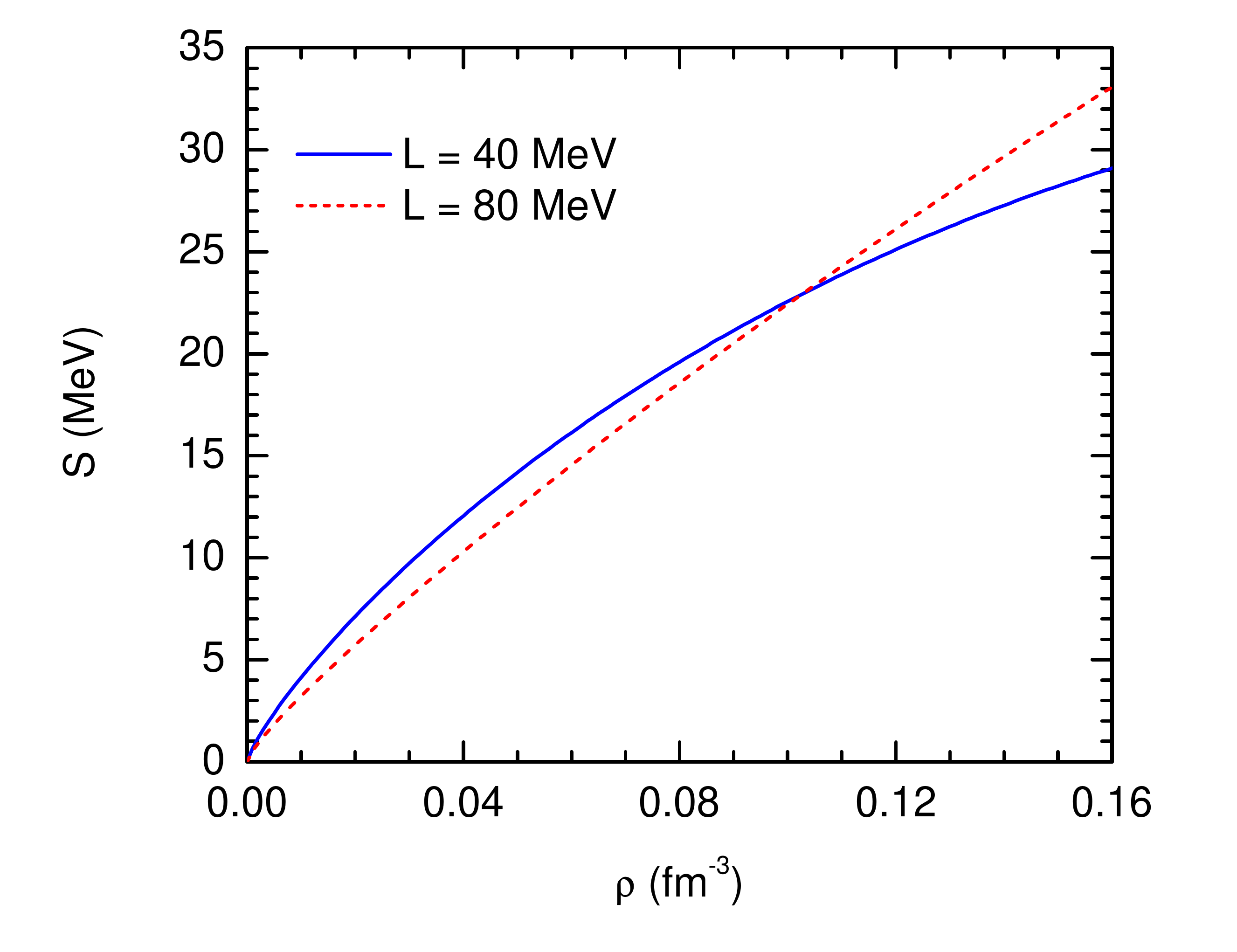}
\caption{(Color online) Density dependence of the nuclear symmetry
energy for the two models discussed in the text.} \label{Fig01}
\end{figure}
In Fig.\,\ref{Fig01} we display the resulting density dependence of
the nuclear symmetry energy for these two interactions. The large
magnitude of the density slope $L$ ensures that at sub-saturation
densities pertaining to the crust of neutron stars the nuclear
symmetry energy acquires smaller values. Thus for large $L$ it
becomes energetically favorable for the system to become more
neutron-rich at these densities. For the same reason, the proton
fraction $Y_{\rm p}$ in the system increases, when $L$ is
small (soft symmetry energy).

\subsection{Grid Spacing and Accuracy Considerations}

As noted above, in \verb"Sky3D"  the wave functions and fields are
defined on a three-dimensional regular Cartesian
grid\,\cite{Maruhn:2013mpa}. In particular, in calculating the
values of Table\,\ref{Table2} we used a cubic box with size $a = 24$
fm and grid spacing of $\Delta x = 1.00$ fm in each direction. As
shown in Ref.\,\cite{Sagert:2015rra}, changing the box size to
larger values does not significantly change the total energies of
the ground state. In fact, doubling the box size can add an
additional energy of only less than 0.012\%. On the other hand, the
choice of the physical spacing between the grid points can be more
important, especially when the grid spacing is larger than $\Delta x
= 1.00$ fm. The calculations presented in Table\,\ref{Table2} uses a
very fine grid of $\Delta x = 0.25$ fm. However this comes at a
significant cost on computational time. For accurate results in
finite nuclei calculations a typical value was suggested to be taken
as $\Delta x \approx 0.75$ fm. Indeed, when we used $\Delta x= 1.00$
fm, the error is less than 0.013\%, whereas the calculation speeds
up by about 100 times. Ideally, the computation time is expected to
scale as $n^3$, where $n$ is the total number of the grid points in
one direction, $n = a/\Delta x$. In Ref.\,\cite{Newton:2007}, it was
shown that the maximum grid spacing for nuclear pasta calculations
may be taken as large as $1.30$ fm. Notice that the simulation
runtime also depends on the number of nucleons $A$, and scales
approximately as $\sim A^2$ due to diagonalization of the
Hartree-Fock hamiltonian. Thus for a fixed average baryon density
and fixed physical spacing between the grid points, doubling the
simulation volume (\textit{i.e} $A \rightarrow 2 A$) makes the
simulation runtime approximately $8$ times longer.

\begin{figure}[tbh]
\vspace{-0.05in}
\includegraphics[width=1.0\columnwidth,angle=0]{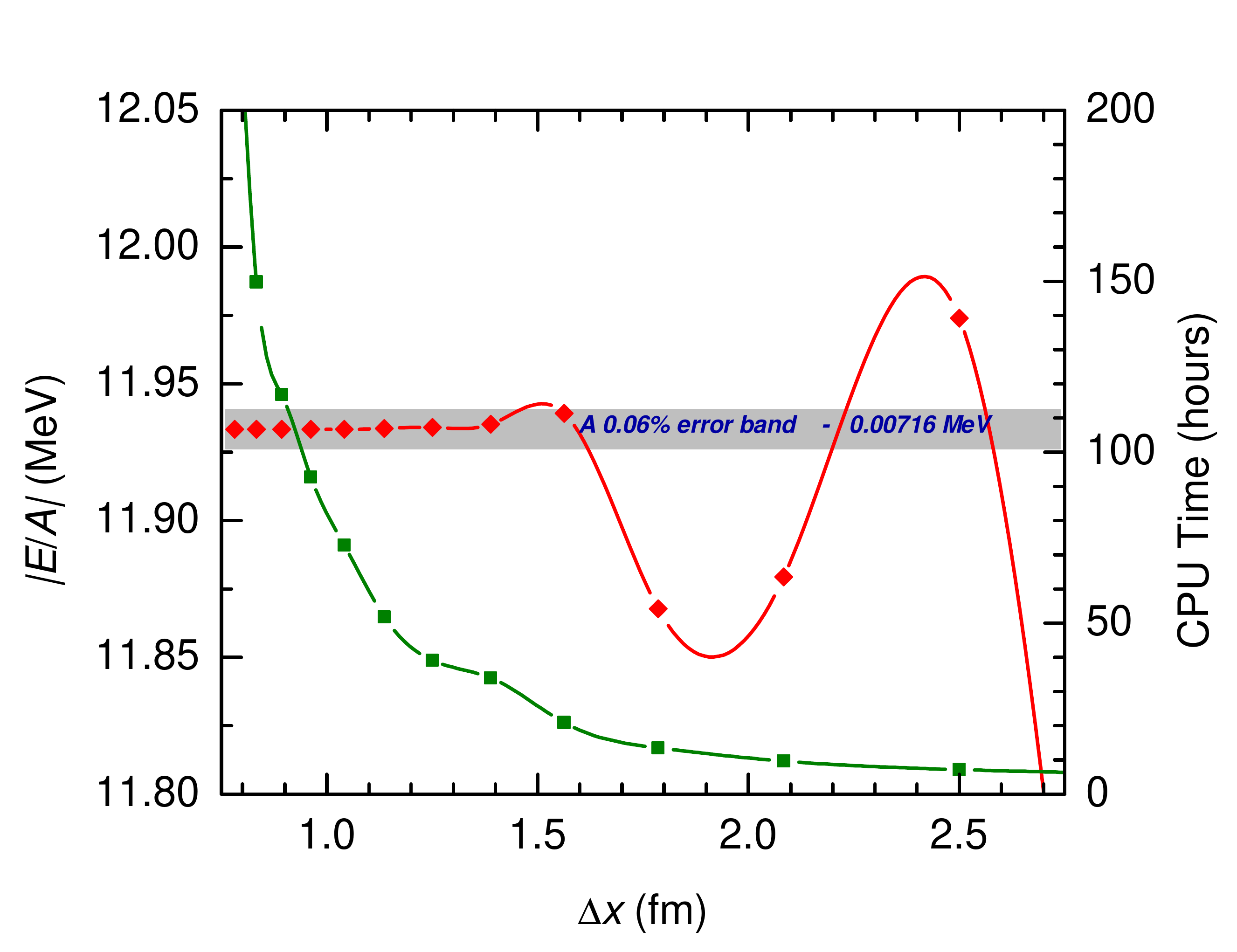}
\caption{(Color online) The absolute value of the binding energy per
nucleon and the simulation runtime as a function of the grid spacing
$\Delta x$ for a system of $A=800$ nucleons at the average baryon
density of $\rho = 0.0512$ fm$^{-3}$ and proton fraction of $Y_{\rm
p} = 0.4$.} \label{Fig02}
\end{figure}

With the aim to maximize the volume of the nuclear pasta systems, we
explored the optimal value of the physical grid spacings that allows
one to perform nuclear pasta simulations without the loss of
accuracy in energies. Notice that nuclear pasta phases are expected
to be sensitive to the binding energy differences of as small as
$0.01$ MeV per nucleon. In Fig.\,\ref{Fig02} we display the absolute
values of the binding energy per nucleon for a system with $A = 800$
nucleons as a function of the grid spacing. We initialize our system
with all nucleons randomly distributed within the box, and the
initial single-particle wave functions are given as a sum of 27
three-dimensional Gaussians with widths of $\sigma = 2.5$ fm that
are centered at the nucleon coordinates with their closest images
formed due periodic boundary conditions. As evident from
Fig.\,\ref{Fig02}---and as far as the binding energies are
concerned---the accuracy of the results are maintained within 0.06\%
for grid spacings of as large as $\Delta x = 1.50$ fm. Moreover,
while the corresponding simulation runtime gets significantly
reduced, an appreciable speed up in the convergence is not observed
beyond $\Delta x
>1.50$ fm. And since the number of grid points must be chosen as even numbers to preserve the
reflection symmetry, we ensure that our grid spacings are chosen as
large as possible but not larger than $\Delta x = 1.50$ fm in our
pasta calculations.

In \verb"Sky3D" the coupled
mean-field equations are solved iteratively. The wave functions are
iterated with a gradient step method which is accelerated by the
kinetic-energy damping (see Ref.\,\cite{Maruhn:2013mpa} for
details):
\begin{equation*}
\psi_{\alpha}^{(n+1)} = \mathcal{O} \left\{\psi_{\alpha}^{(n)} -
\frac{\delta}{\hat{T} + T_0} \left(\hat{h}^{(n)} - \langle
\psi_{\alpha}^{(n)}| \hat{h}^{(n)} |  \psi_{\alpha}^{(n)} \rangle
\right) \psi_{\alpha}^{(n)} \right\} \ ,
\end{equation*}
where $\hat{T} = \hat{p}^2/2m$ is the operator of kinetic energy,
$\mathcal{O}$ means orthonormalization of the whole set of new wave
functions, $\hat{h}$ is single-particle hamiltonian, and the upper
index indicates the iteration number. The damped gradient step has
two numerical parameters, the step size $\delta$ and the damping
regulator $T_0$. Ref.\,\cite{Maruhn:2013mpa} suggests a value of
$\delta = 0.1 \ldots 0.8$ and $T_0 = 100$ MeV should be optimal.
Larger values of $\delta$ yield faster iteration, but can run more
easily into pathological conditions.

In an effort to optimize our simulation we introduced a variable
step size that starts with an initial $\delta =0.2$ and is
systematically increased by a factor of $1.005$ if the new
single-particle energies are smaller than the one from the previous
iteration, otherwise it is decreased by a factor of $1.250$. This
ensures in average an about three times faster convergence than when
a constant $\delta$ is assumed.

To avoid getting trapped in a metastable state we run our simulations
very long and have chosen our convergence criterion to be $\Delta
\epsilon_{\rm tot} =  \epsilon_{\rm tot}^{(m)} - \epsilon_{\rm
tot}^{(n)} < - 10^{-4}$ MeV, where $\epsilon_{\rm tot}$ is total
energy per nucleon at a given iteration, and $m = n + 10,000$. The
total energy of the ground state is then found as $\epsilon_{\rm
g.s.} = \epsilon_{\rm tot}^{(m)}$.

\subsection{Finite Size Effects and the Minimum Number of Nucleons}
\begin{figure}[p]
 \begin{center}
  \includegraphics[width=0.26\textwidth]{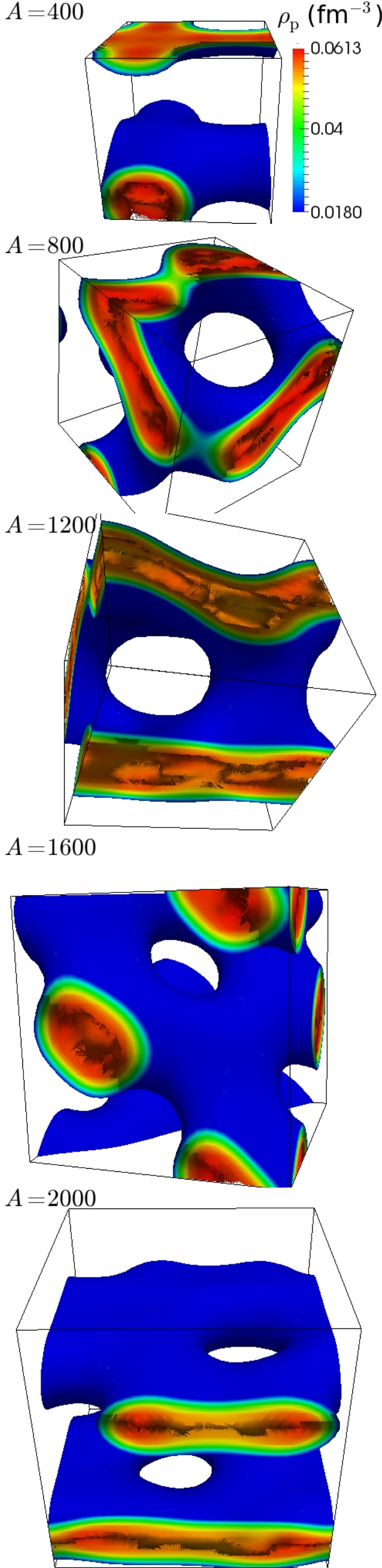}
 \end{center}
   \caption{(Color online)
  The appearance of the nuclear waffle phase at $Y_{\rm p} = 0.40$, $\rho = 0.05$ fm$^{-3}$ for different volume sizes containing
  $A =$ 400, 800, 1200, 1600 and 2000 nucleons from top to the bottom, respectively. The sides of the cubic volumes
  correspondingly are equal to $20$ fm, $25.2$ fm, $28.8$ fm, $31.7$ fm, $34.2$ fm. The blue color represents isosurface proton densities of
  $\rho_{\rm p} = 0.9  \left(Y_{\rm p} \rho \right)$ and the red color represents the region with the highest proton density $\rho_{\rm p}^{\rm max}$
  within the pasta structure, where $\rho$ is the average nucleon density. This figure and all other similar figures
  throughout the paper are  generated using the ParaView software\,\cite{Paraview}.}
 \label{Fig03}
\end{figure}

  \begin{table*}[t]
  \begin{tabular}{|r|c|c|c|c|c|c|c|c|c|c|}
    \hline
    $A$    & $E_{\rm tot}$ & $E_{\rm kin}$ & $E_0$      & $E_1$     & $E_2$    & $E_3$     & $E_{\rm ls}$  & $E_{\rm C}$ & $\rho_{\rm tot}^{\rm max}$ & $\rho_{\rm p}^{\rm max}$  \\
    \hline
    \hline
    $400$  & $-11.8565$    & $18.9389$     & $-94.0201$ & $-0.3275$ & $0.6632$ & $62.3350$ & $-0.2240$     & $0.7780$    & $0.1486$                   & $0.0612$  \\
    $800$  & $-11.8164$    & $18.7521$     & $-92.4597$ & $-0.3081$ & $0.7419$ & $61.1158$ & $-0.2592$     & $0.6008$    & $0.1491$                   & $0.0613$  \\
    $1200$ & $-11.8320$    & $18.9109$     & $-93.6282$ & $-0.3226$ & $0.6640$ & $61.9873$ & $-0.2340$     & $0.7905$    & $0.1497$                   & $0.0619$  \\
    $1600$ & $-11.8609$    & $19.0713$     & $-94.8028$ & $-0.3240$ & $0.7245$ & $63.0520$ & $-0.2606$     & $0.6787$    & $0.1515$                   & $0.0630$  \\
    $2000$ & $-11.8520$    & $18.8880$     & $-93.4096$ & $-0.3150$ & $0.7394$ & $61.8989$ & $-0.2639$     & $0.6103$    & $0.1529$                   & $0.0645$  \\
   \hline
  \end{tabular}
 \caption{Various contributions to the total energy of the system are given in units of MeV for nuclear pasta configurations
          with different number of nucleons, $A$, at a fixed average baryon density of $\rho = 0.05$ fm$^{-3}$
          and proton fraction of $Y_{\rm p} = 0.40$. To make a meaningful comparison
          between these systems, energies per nucleon are presented only. Also the maximum local total density $\rho_{\rm tot}^{\rm max}$ is given,
          as well as the maximum local proton density, $\rho_{\rm p}^{\rm max}$, within the pasta structure in units of fm$^{-3}$.}
 \label{Table3}
 \end{table*}
Having settled on the optimum choice of the grid spacing, in this
subsection we explore the role of the finite size effects on the
energetics and geometries of the nuclear pasta. In Fig.\,\ref{Fig03}
we plotted the isosurface of proton densities for systems with $A =
$ $400$, $800$, $1200$, $1600$ and $2000$ nucleons, respectively, at
a fixed average baryon density of $\rho = 0.05$ fm$^{-3}$ and proton
fraction of $Y_{\rm p} = 0.40$. In Table\,\ref{Table3} we present
the corresponding energetics and maximum local densities.

It turns out, that all of these systems are energetically very close
to one another with accuracy of less than 0.0445 MeV in the binding
energy per nucleon. Nevertheless, as depicted in Fig.\,\ref{Fig03}
the corresponding pasta phases assume a seemingly different shape
for each case. Considering that these systems obey periodic boundary
conditions it is not difficult to see that most of them are in the
\textit{nuclear waffle} state with the exception of $A=800$ and $A =
1200$, where there are additional 3D
connections\,\cite{Schneider:2014lia}. The existence of nuclear
waffles as perforated plates was observed by
Ref.\,\cite{Schneider:2014lia} using MD simulations and also in
\cite{Schuetrumpf:2012cj}, where it was denoted as the rod(2) shape.
This phase is expected to lie in the transition between a phase made
up of elongated cylindrical nuclei and a phase formed of a stack of
parallel flat plates. Recently it was shown\,\cite{Sagert:2015rra}
that if the simulation is initialized from the single-particle
wavefunctions constructed from a converged MD simulation, the waffle
state remains stable even when quantum mechanical effects are
considered. Even starting from a completely random initial
configuration we confirm that the waffle state is a true stable
nuclear pasta configuration in agreement with the results obtained
by Ref.\,\cite{Schuetrumpf:2012cj,
Schneider:2014lia}. 
Looking more closely at the individual energy components as given in
Table\,\ref{Table3}, we realize that the highest percentage error
comes from the Coulomb energy contribution. This is because the
Coulomb force has long-range interaction and can extend much beyond
the boundaries of smaller boxes. The individual energy terms from
Skyrme force have also larger percentage errors as opposed to the
total energy. This is primarily due to the fact that the ground
state is, by definition, obtained by minimizing the total energy.
Therefore individual terms can have different values stemming from
the competition between nuclear and electric forces and as a result
of their overall effort to minimize the ground state energies. Thus,
although the final ground state energies are close to one another,
the final shape of the nuclear pasta depends on the system size as a
result of such competition. Following Fig.\,\ref{Fig03} where we
obtained at least two pasta structures for a system with an average
baryon density of $\rho = 0.05$ fm$^{-3}$, in the next part of our
discussions we assume systems containing $A = 2000$ nucleons.

\section{Results}
\label{Results}

\subsection{Neutron Star Crust: $Y_{\rm p} = 0.05$}
\begin{figure}[tb]
 \begin{center}
\includegraphics[width=1.0\columnwidth,angle=0]{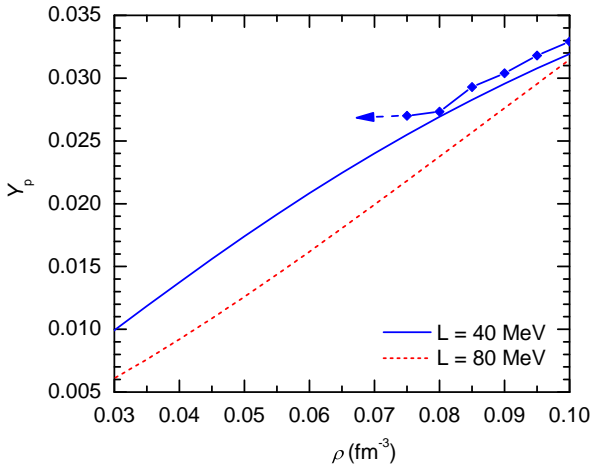}
 \end{center}
   \caption{(Color online)
 Proton fractions as a function of baryon density for a uniform neutron-star matter in two models discussed in the text.
 Also shown is the proton fractions at few fixed average baryon densities obtained directly from nuclear pasta simulations (diamonds).}
 \label{Fig04}
\end{figure}
Every simulation described here has $A = 2000$ nucleons. These
nucleons are initially randomly positioned within a cubic box with
sides $a = \sqrt[3]{A/\rho}$ and corresponding initial
single-particle wave functions are constructed by folding Gaussians
over each nucleon. We present and discuss our results for a fixed
proton fraction of $Y_{\rm p} = 0.05$. This condition mimics the
matter content in the neutron star crust. For a proper description
of the neutron-star matter, one must obtain proton fractions
self-consistently by using the condition of chemical equilibrium:
\begin{equation}
\label{BetaEq} \mu_{\rm n} = \mu_{\rm p} + \mu_{\rm e} \ ,
\end{equation}
where $\mu_{\rm q}$ is the chemical potential of species $\rm q = n,
p, e$ for neutrons, protons and electrons, respectively. Assuming a
uniform nuclear matter in beta-equilibrium we find that both
interactions predict proton fractions to be less than 5\% at
densities of $0.03\, {\rm fm}^{-3} < \rho < 0.10\, {\rm fm}^{-3}$
where the emergence of nuclear pasta is expected, see
Fig.\,\ref{Fig04}. In this figure we also display proton fractions
at a few fixed baryon densities which were obtained directly from
the nuclear pasta simulations. For this we fixed the proton number
at $Z=14$ and varied the neutron number, $N = A-Z$, in search for
the value of $A$ that satisfies the condition (\ref{BetaEq}). Notice
that this search is quite exhausting as far as the simulation
computing times are concerned. Moreover, for realistic results one
must choose proton numbers to be $Z \gg 14$. We reserve to carry out
such simulations in the future. However, we would like to point out
that whereas at densities close to saturation the proton fractions
closely match that obtained from a uniform matter distribution, at
lower sub-saturation densities the realistic proton fractions can be
larger due clustering effects as hinted by the left arrow in
Fig.\,\ref{Fig04}. The question of whether exotic structure phases
can develop in a proton-deficient environment was critically
analyzed by Ref.\,\cite{Piekarewicz:2011qc}. In particular, they
found an interesting behavior displayed in the structure factor
$S(q)$ that could be indicative of significant structural changes in
the system. Nevertheless, it was concluded that no clear evidence
exists either in favor or against the formation of the nuclear pasta
at the neutron crust. To our knowledge, no other full quantum
numerical simulations have been carried out with proton fractions
less than $Y_{\rm p} = 0.1$.

\begin{figure*}[h]
 \begin{center}
  \includegraphics[width=0.47\textwidth]{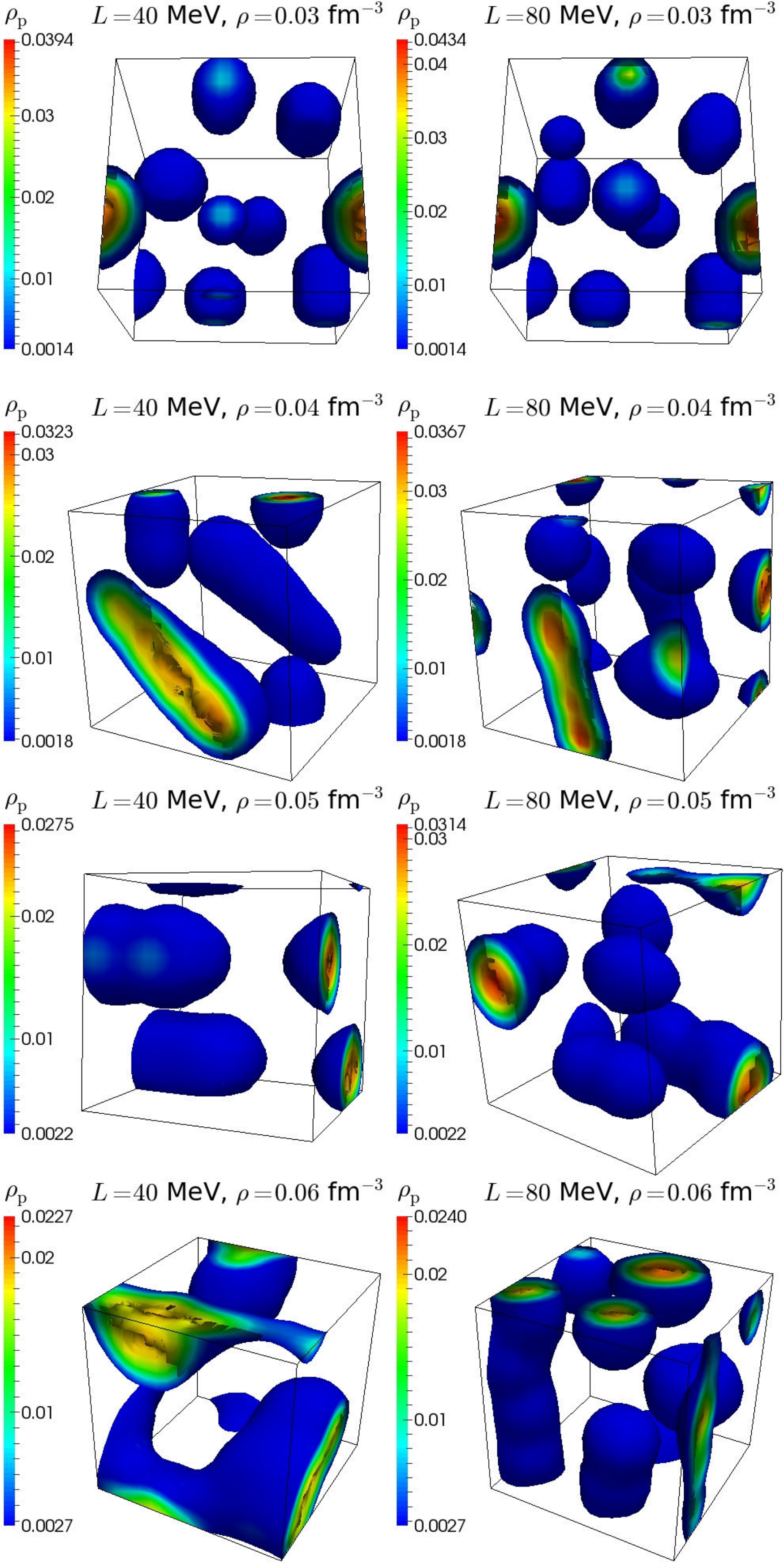}
        \hspace{0.1cm}
  \includegraphics[width=0.47\textwidth]{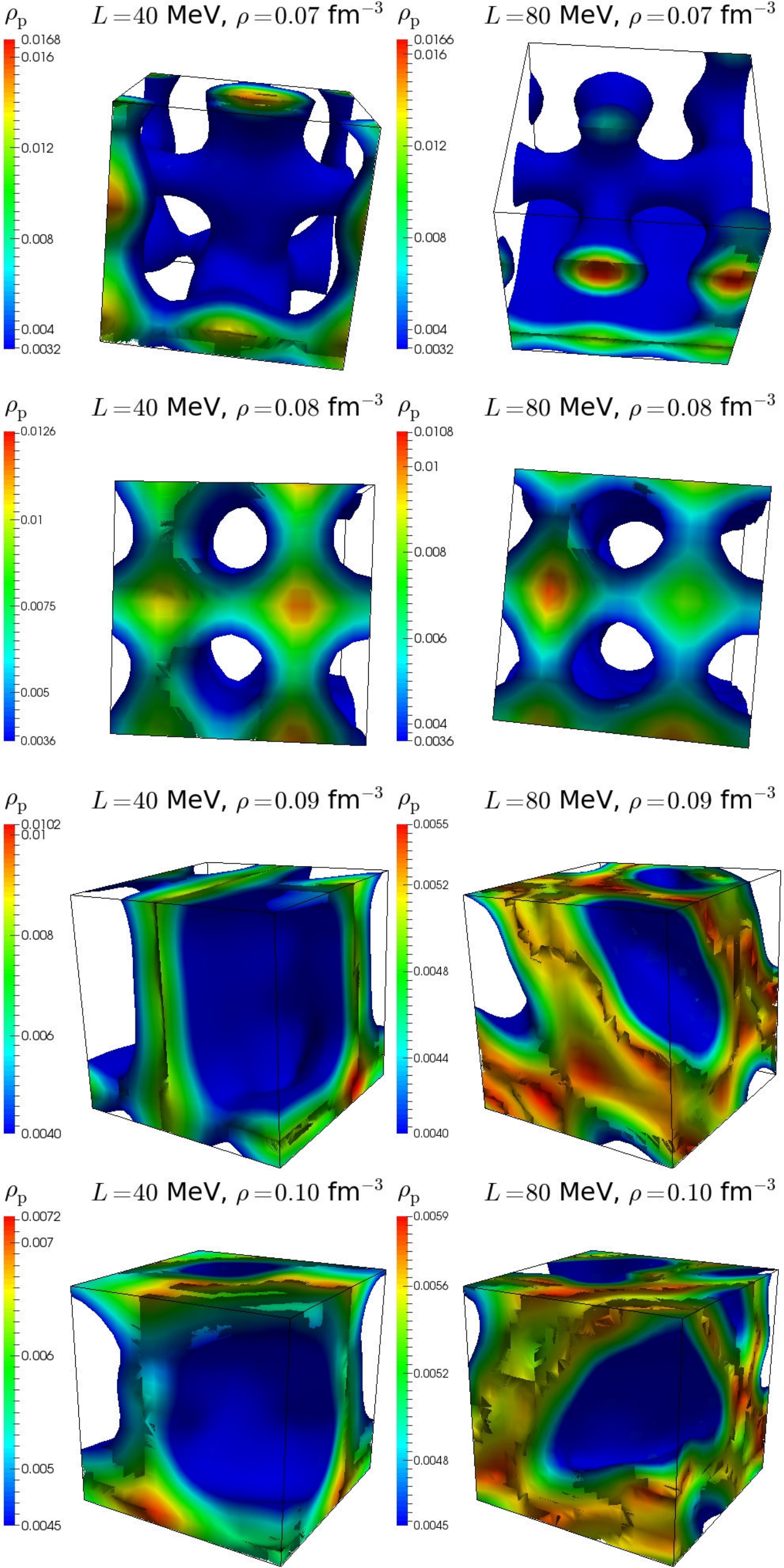}
  \caption{(Color online)
  Isosurface of proton densities are plotted for the two model discussed in the text over the range of baryon densities at a
  fixed proton fraction of $Y_{\rm p} = 0.05$. The total number of nucleons are fixed at $A = 2000$ and the side of the cubic
  box varies from $40.55$ fm down to $27.14$ fm, corresponding to average baryon densities of $0.03 \leq \rho \leq 0.10$ fm$^{-3}$, respectively.}
 \label{Fig05}
 \end{center}
\end{figure*}
  \begin{table}[t]
  \begin{tabular}{|c|l|c|c|c|c|c|}
    \hline
    $\rho$  & Model                     & $E_{\rm tot}$ (MeV) & $\rho_{\rm tot}^{\rm min}$ & $\rho_{\rm tot}^{\rm max}$ & $N_{\rm f}$            & $Y_{\rm p}^{\star}$ (\%)  \\
    \hline
    \hline
    $0.03$  & \mbox{\tt UNEDF1}            & $1.731$             & $0.0217$                   & $0.1437$                   & $\phantom{-}786$       & ~$8.24$ \\
            & \mbox{\tt UNEDF1}$^{\star}$   & $0.481$             & $0.0225$                   & $0.1376$                   & $\phantom{-}508$       & ~$6.70$ \\
    \hline
    $0.04$  & \mbox{\tt UNEDF1}            & $2.118$             & $0.0300$                   & $0.1316$                   & $\phantom{-}788$       & ~$8.25$ \\
            & \mbox{\tt UNEDF1}$^{\star}$   & $0.801$             & $0.0312$                   & $0.1275$                   & $\phantom{-}581$       & ~$7.05$ \\
    \hline
    $0.05$  & \mbox{\tt UNEDF1}             & $2.522$             & $0.0369$                   & $0.1285$                   & $\phantom{-}810$       & ~$8.40$ \\
            & \mbox{\tt UNEDF1}$^{\star}$   & $1.212$             & $0.0405$                   & $0.1232$                   & $\phantom{-}674$       & ~$7.54$ \\
    \hline
    $0.06$  & \mbox{\tt UNEDF1}            & $2.937$             & $0.0456$                   & $0.1222$                   & $\phantom{-}848$       & ~$8.68$ \\
            & \mbox{\tt UNEDF1}$^{\star}$   & $1.715$             & $0.0490$                   & $0.1132$                   & $\phantom{-}770$       & ~$8.13$ \\
    \hline
    $0.07$  & \mbox{\tt UNEDF1}             & $3.356$             & $0.0537$                   & $0.1111$                   & $\phantom{-}874$       & ~$8.88$ \\
            & \mbox{\tt UNEDF1}$^{\star}$   & $2.292$             & $0.0594$                   & $0.1034$                   & $\phantom{-}902$       & ~$9.11$ \\
    \hline
    $0.08$  & \mbox{\tt UNEDF1}             & $3.778$             & $0.0631$                   & $0.1061$                   & $\phantom{-}926$       & ~$9.31$ \\
            & \mbox{\tt UNEDF1}$^{\star}$   & $2.962$             & $0.0711$                   & $0.0975$                   & $\phantom{-}978$       & ~$9.78$ \\
    \hline
    $0.09$  & \mbox{\tt UNEDF1}             & $4.229$             & $0.0753$                   & $0.1079$                   & $\phantom{-}978$       & ~$9.78$ \\
            & \mbox{\tt UNEDF1}$^{\star}$   & $3.732$             & $0.0828$                   & $0.0924$                   & $1122$                 & $11.39$ \\
    \hline
    $0.10$  & \mbox{\tt UNEDF1}            & $4.716$             & $0.0870$                   & $0.1071$                   & $\phantom{-}978$       & ~$9.78$ \\
            & \mbox{\tt UNEDF1}$^{\star}$   & $4.601$             & $0.0939$                   & $0.1021$                   & $1218$                 & $12.79$ \\
   \hline
  \end{tabular}
 \caption{Some bulk properties of nuclear pasta with average proton fraction of
          $Y_{\rm p} = 0.05$. Here $N_{\rm f}$ represents the number of free neutrons
          and $Y_{\rm p}^{\star} = Z/(A-N_{\rm f})$ is defined as the
          effective proton fraction of the pasta structure. All densities are given in units of fm$^{-3}$.}
 \label{Table4}
 \end{table}
In Fig.\,\ref{Fig05} we plot the isosurface of proton densities for
models with both soft, $L =40$ MeV, and stiff, $L =80$ MeV symmetry
energies. At the lowest density of $\rho = 0.03$ fm$^{-3}$
considered in our simulations, we observe a combined total of $8$
spherical and deformed nuclei, unequal in size, in both models.
Their location is randomly distributed within the box and do not
form a lattice structure of any kind. Notice that such density
already corresponds to a deeper layer of the inner crust. The
transition from the outer crust to the inner crust is predicted to
occur at about $\rho
> 0.00024$ fm$^{-3}$\,\cite{Baym:1971pw, RocaMaza:2008cg}.
Whereas at the top layer of the inner crust one expects a Coulomb
crystal of neutron-rich nuclei immersed in a uniform electron gas
and a dilute neutron vapor\,\cite{Piekarewicz:2011qc}, at $\rho =
0.03$ fm$^{-3}$ the neutron vapor becomes much denser, and the
crystalline structure is already destroyed. These so-called
\textit{gnocchi} phase could be said to form a liquid-like (or
amorphous) structure with an approximate average charge of $\langle
Z \rangle \approx 12.5$. This likely is because the system is not
equilibrated. These nuclei are well separated from one another, and
their sizes and shapes are mostly dictated by the Coulomb repulsion
between protons and the surface energy of the system, which are
almost identical in both cases. The corresponding total energies per
nucleon in these two models are surprisingly different (see
Table\,\ref{Table4}). This difference primarily comes from the
zero-range term $E_0$ and density dependent term $E_3$, whose values
strongly depend on $x_0$ and $x_3$ Skyrme parameters,
respectively\,\cite{Chabanat:1997qh}. Physically, a large slope
parameter $L$ means that the symmetry energy at low densities is
small, thus nuclei can easily become neutron-rich. On the other
hand, the symmetry energy at $\rho = 0.03$ fm$^{-3}$ is larger for a
model with small value of $L$, thus it becomes energetically
favorable for the system to maintain larger proton fractions. We
further examined the single-particle energies and have found that
the number of free neutrons, $N_{\rm f}$, identified as the number
of neutrons with positive single-particle energies, is indeed
smaller for $L = 80$ MeV than $L=40$ MeV, see Table\,\ref{Table4}.
Thus the system became effectively neutron-rich with an effective
proton fraction $Y_{\rm p}^{\star} = Z/(A-N_{\rm f})$ being smaller
in the former.

\begin{figure}[tb]
 \begin{center}
  \includegraphics[width=0.5\textwidth]{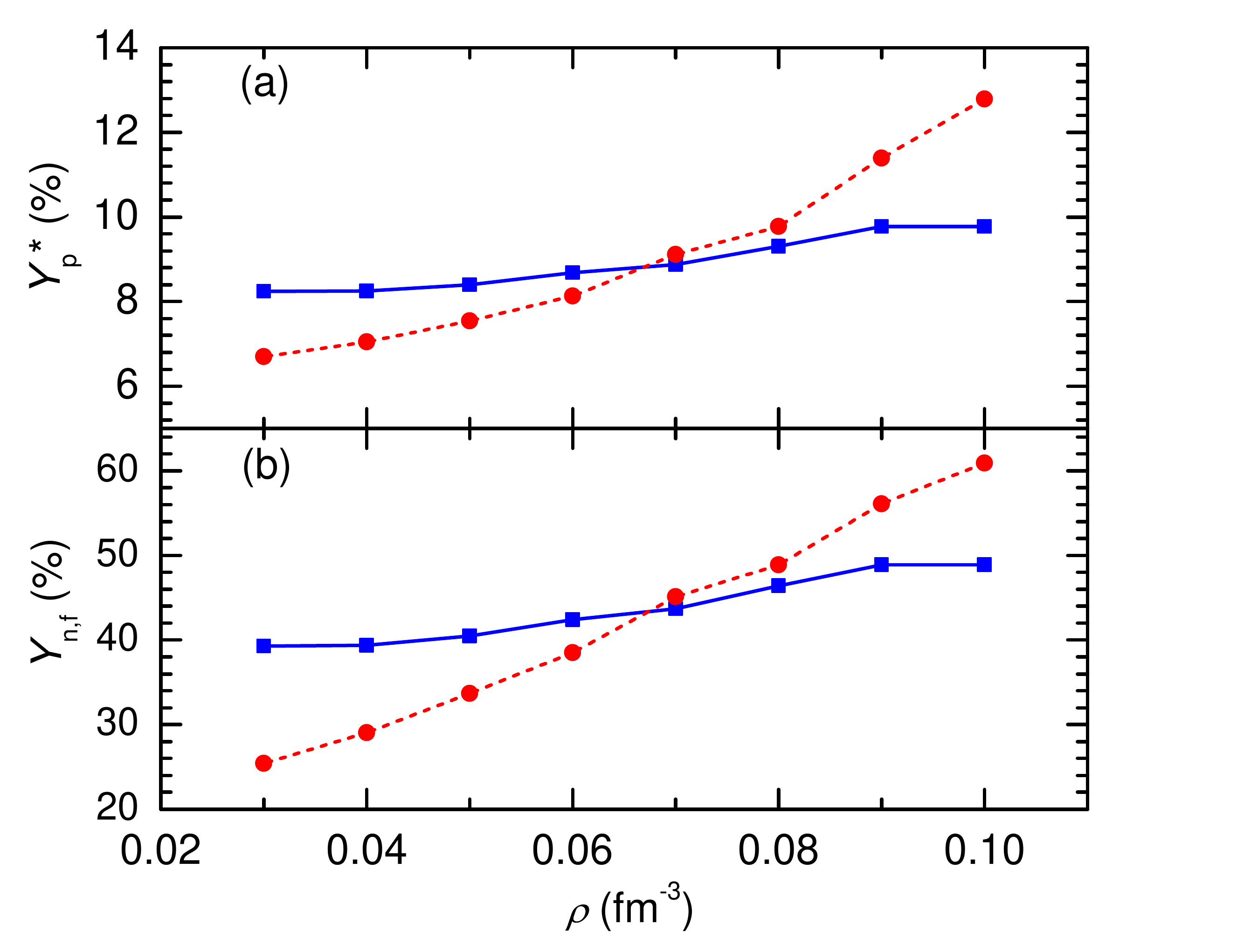}
 \end{center}
   \caption{(Color online)
   Effective proton fractions $Y_{\rm p}^{\star}$ (top panel) and free neutron fractions
   $Y_{\rm n, f}$ (bottom panel) are plotted
   as a function of total average baryon density for the two models discussed in the text.}
 \label{Fig06}
\end{figure}
As the average baryon density increases to $\rho = 0.04$ fm$^{-3}$,
the nuclei come closer, get fused and merge into super-elongated
nuclei of rod-like structure, see Fig.\,\ref{Fig05}. Whereas all of
the 8 nuclei got merged to 3 rod-like structures in the model with
the soft symmetry energy, only 1 rod-like structure and 5 nuclei are
observed in the model with $L=80$ MeV, thus harboring a coexistence
of two structures: spherical nuclei and super-elongated nuclei of
rod-like behavior. Note again that this result is likely due to the
system being not equilibrated. At even higher density of $\rho =
0.05$ fm$^{-3}$, the former now has 2 rod-like structures only,
whereas the latter has 3 rod-like structures and 2 nuclei within the
simulation box. The corresponding effective proton fractions rise in
both models, meaning there are more free neutrons in the system now
(see Table\,\ref{Table4}).
Since the symmetry energy rises faster as a
function of density in the model with $L = 80$ MeV, the effective
proton fraction also gets boosted further as evidenced by the
results shown on Table\,\ref{Table4} and displayed in
Fig.\,\ref{Fig06}.

At $\rho = 0.06$ fm$^{-3}$, in \mbox{\tt UNEDF1}, the rod-like
structures now start getting fused in the perpendicular direction.
As density is increased to $\rho = 0.07$ fm$^{-3}$ rods get further
fused and the system is comprised of a continuous crest-like
structure (recall that the system is periodic). On the other hand,
at $\rho = 0.06$ fm$^{-3}$, the phase co-existence between rods and
nuclei continue to exist in \mbox{\tt UNEDF1}$^{\star}$, whereas at
$\rho = 0.07$ fm$^{-3}$, we observe a combination of
$P$-surface\,\cite{Schuetrumpf:2015nza} and a flat plate, also known
as the \textit{lasagna} phase. This means that pure rod-like
structures in models with the stiff symmetry energy can only exist
within a very narrow region of densities. Correspondingly, only a
very thin layer of such pasta can exist in the neutron star crust.

At $\rho = 0.08$ fm$^{-3}$ in both systems we observe hollow-tubes,
also known as the \textit{bucatini} phase. More neutrons become free
than bound. The corresponding effective proton fractions, $Y_{\rm
p}^{\star}$, and free neutron fractions, $Y_{\rm n, f} = N_{\rm
f}/A$, as a function of density are plotted in the left and right
panels of Fig.\,\ref{Fig06}.

Finally, we observe spherical bubbles, also known as the
\textit{Swiss cheese} phase, at densities of $\rho = 0.09$ fm$^{-3}$
and $\rho = 0.10$ fm$^{-3}$. The sizes of spherical bubbles get
smaller as the density increases and also depend on the two models
considered above.

\begin{figure}[tb]
 \begin{center}
  \includegraphics[width=0.5\textwidth]{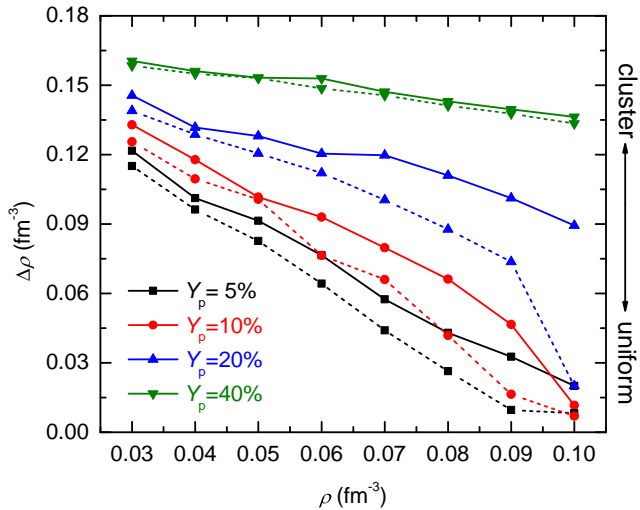}
 \end{center}
   \caption{(Color online)
   Density contrast $\Delta \rho$ within the pasta simulation box as a function of
   the average baryon density $\rho$ for various proton fractions $Y_{\rm p}$
   for models with $L=40$ MeV (solid) and $L=80$ MeV (dashed). System is considered as
   uniform when $\Delta \rho = 0$, otherwise it is pasta-rich.}
 \label{Fig07}
\end{figure}
Particularly interesting is to observe the density contrast, $\Delta
\rho = \rho_{\rm tot}^{\rm max} - \rho_{\rm tot}^{\rm min}$, within
the nuclear pasta systems described above. Here $\rho_{\rm tot}^{\rm
max}$ and $\rho_{\rm tot}^{\rm min}$ are the maximum and the minimum
local baryon densities within the simulation volume. In particular,
$\rho_{\rm tot}^{\rm max}$ is the baryon density at the central
regions of pasta structures, whereas $\rho_{\rm tot}^{\rm min}$ is
the baryon density of the free neutron gas. The larger value of
$\Delta \rho$ suggests that the system organized itself into complex
clusters, whereas $\Delta \rho = 0$ means the system is uniform.
Considering Table\,\ref{Table4} and Fig.\,\ref{Fig07} we see that a
soft symmetry energy exhibits a pasta-rich system throughout the
neutron star crust, whereas the clustered matter transforms quickly
into the uniform matter when the symmetry energy is stiff (see
Table\,\ref{Table4}).

\subsection{Proto-Neutron Stars and Matter in Supernova}
In cold neutron stars proton fractions of larger than $Y_{\rm p} >
0.05$ can only occur at high densities and very low densities. At
high densities pertaining to the core of the star the matter is
uniform and no nuclear pasta phase is therefore expected. Similarly,
at very low densities applicable to the outer crust, nucleons bind
into nuclei that are then segregated in a crystal lattice. However,
the low-density regions that contain proton fractions between $0.10
< Y_{\rm p} < 0.40$ can be present in dense proto-neutron stars
(PNS) that are born subsequent to the core-collapse supernova
explosion. The PNS is cooled primarily by neutrino emission which is
driven by neutrino diffusion and convection within the PNS after the
core bounces. It is therefore interesting to understand the role of
the neutrino-matter interaction in the dynamics of the supernova
explosion. The spectrum of neutrinos emerged from neutrino-sphere
can be observed using the current and future terrestrial detectors
as soon as the next galactic or near-galactic supernova goes off.
This spectrum can provide a valuable information about the structure
of the nuclear matter in these regions\,\cite{Horowitz:2016fpa}.

\subsubsection{Systems with $Y_{\rm p} = 0.10$}
\begin{figure*}[h]
 \begin{center}
  \includegraphics[width=0.47\textwidth]{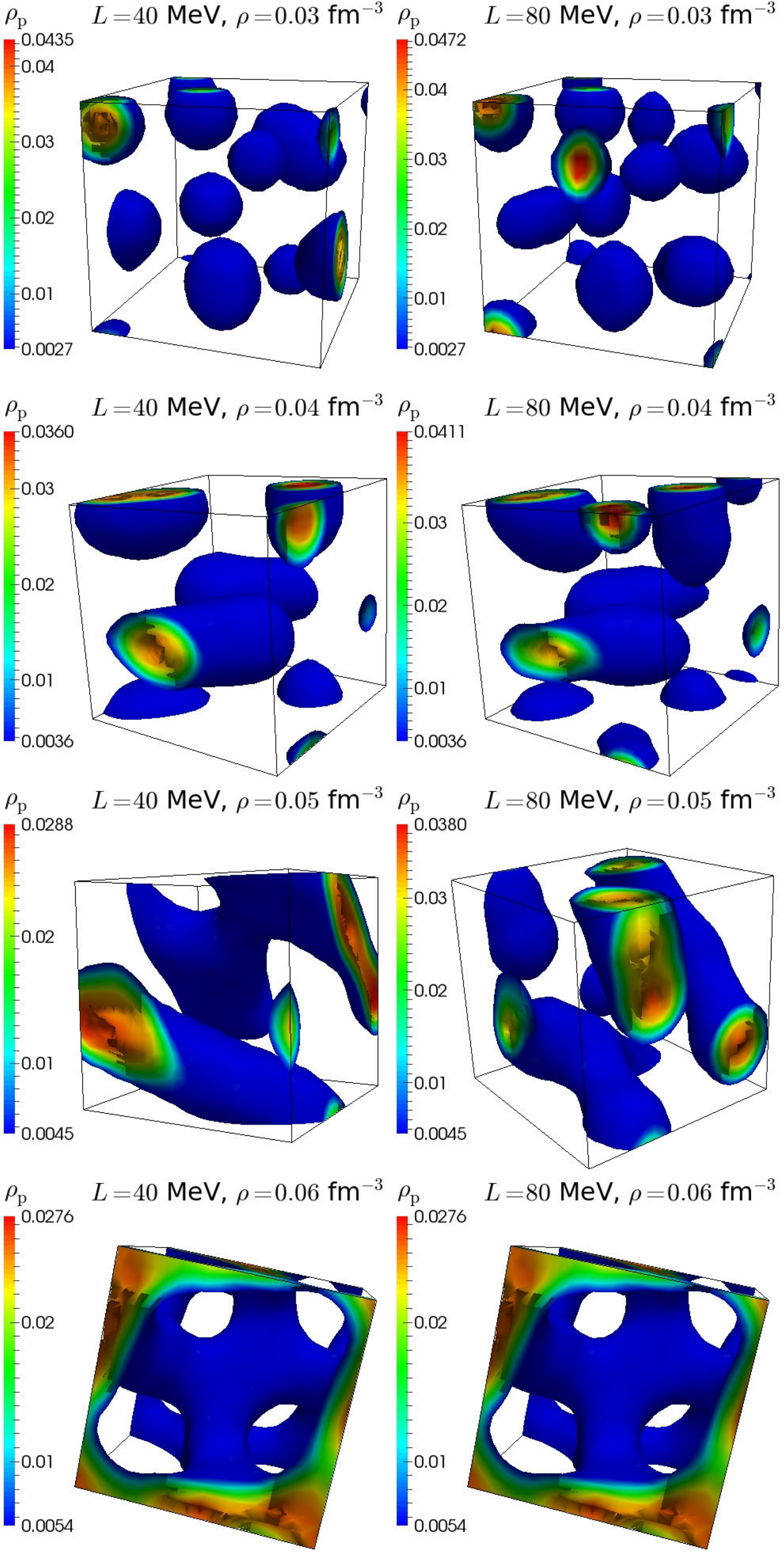}
        \hspace{0.1cm}
  \includegraphics[width=0.47\textwidth]{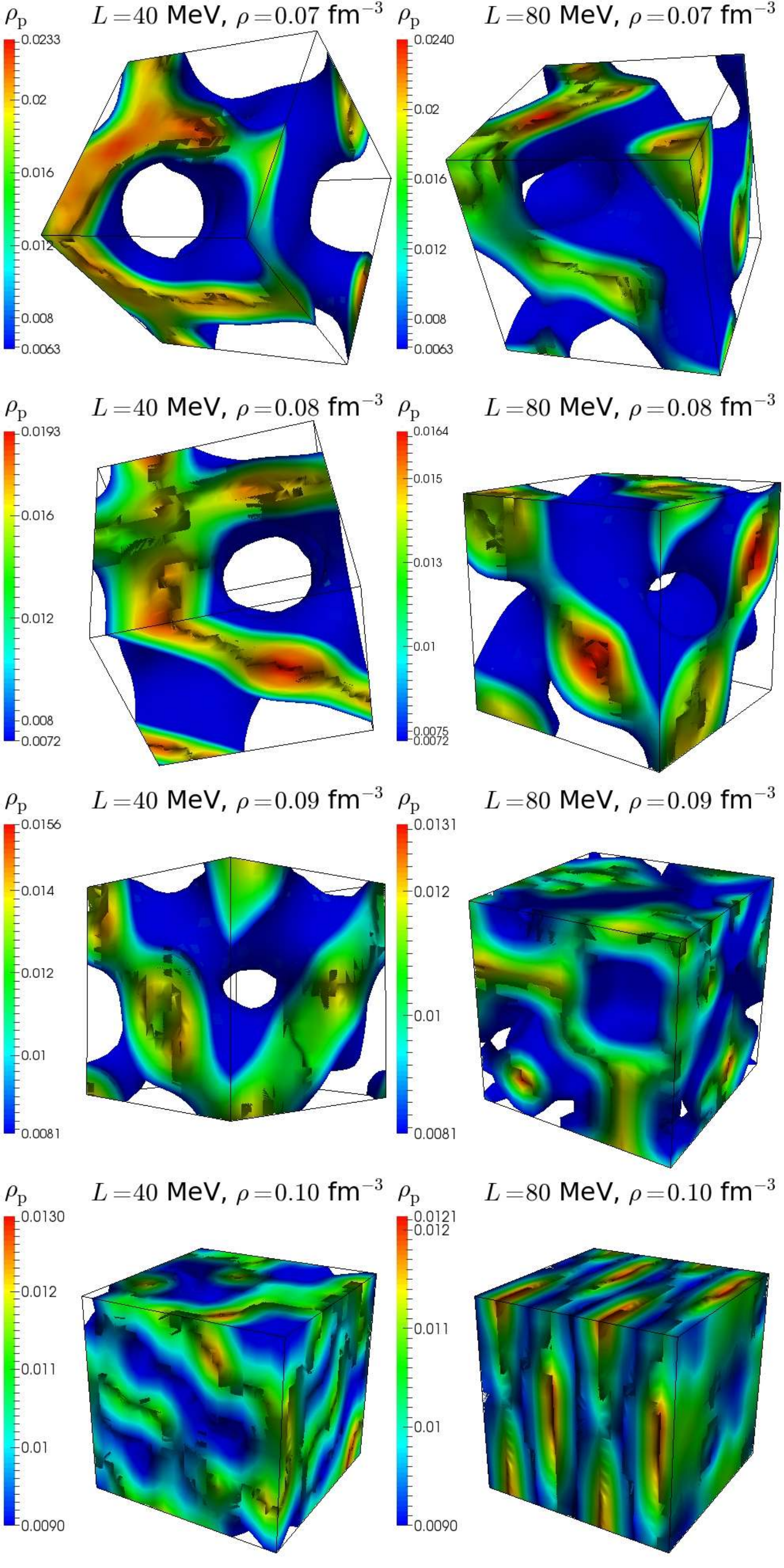}
  \caption{(Color online)
  Isosurface of proton densities are plotted using the same presciription as
  in Fig.\,\ref{Fig05} except now the proton fraction of the system is $Y_{\rm p} = 0.10$.}
 \label{Fig08}
 \end{center}
\end{figure*}
  \begin{table}[t]
  \begin{tabular}{|c|l|c|c|c|c|c|}
    \hline
    $\rho$  & Model                     & $E_{\rm tot}$ (MeV) & $\rho_{\rm tot}^{\rm min}$ & $\rho_{\rm tot}^{\rm max}$ & $N_{\rm f}$ & $Y_{\rm p}^{\star}$ (\%)  \\
    \hline
    \hline
    $0.03$  &\mbox{\tt UNEDF1}             & $-0.668$            & $0.0159$                   & $0.1485$                   & $638$       & $14.68$ \\
            & \mbox{\tt UNEDF1}$^{\star}$   & $-1.537$            & $0.0167$                   & $0.1423$                   & $394$       & $12.45$ \\
    \hline
    $0.04$  & \mbox{\tt UNEDF1}             & $-0.519$            & $0.0206$                   & $0.1382$                   & $630$       & $14.60$ \\
            & \mbox{\tt UNEDF1}$^{\star}$   & $-1.436$            & $0.0239$                   & $0.1334$                   & $438$       & $12.80$ \\
    \hline
    $0.05$  & \mbox{\tt UNEDF1}             & $-0.330$            & $0.0265$                   & $0.1275$                   & $642$       & $14.73$ \\
            & \mbox{\tt UNEDF1}$^{\star}$   & $-1.258$            & $0.0303$                   & $0.1310$                   & $496$       & $13.30$ \\
    \hline
    $0.06$  & \mbox{\tt UNEDF1}             & $-0.137$            & $0.0325$                   & $0.1256$                   & $654$       & $14.86$ \\
            & \mbox{\tt UNEDF1}$^{\star}$   & $-1.018$            & $0.0393$                   & $0.1157$                   & $566$       & $13.95$ \\
    \hline
    $0.07$  & \mbox{\tt UNEDF1}             & $\phantom{-}0.051$  & $0.0399$                   & $0.1197$                   & $666$       & $14.99$ \\
            & \mbox{\tt UNEDF1}$^{\star}$   & $-0.713$            & $0.0466$                   & $0.1126$                   & $644$       & $14.75$ \\
    \hline
    $0.08$  & \mbox{\tt UNEDF1}          & $\phantom{-}0.252$  & $0.0474$                   & $0.1136$                   & $698$       & $15.36$ \\
            &\mbox{\tt UNEDF1}$^{\star}$   & $-0.363$            & $0.0591$                   & $0.1009$                   & $770$       & $16.26$ \\
    \hline
    $0.09$  & \mbox{\tt UNEDF1}          & $\phantom{-}0.457$  & $0.0623$                   & $0.1089$                   & $770$       & $16.26$ \\
            & \mbox{\tt UNEDF1}$^{\star}$   & $\phantom{-}0.068$  & $0.0825$                   & $0.0989$                   & $878$       & $17.83$ \\
    \hline
    $0.10$  & \mbox{\tt UNEDF1}          & $\phantom{-}0.688$  & $0.0956$                   & $0.1073$                   & $830$       & $17.09$ \\
            & \mbox{\tt UNEDF1}$^{\star}$   & $\phantom{-}0.597$  & $0.0975$                   & $0.1045$                   & $926$       & $18.62$ \\
   \hline
  \end{tabular}
 \caption{Some bulk properties of nuclear pasta with an average proton fraction of
          $Y_{\rm p} = 0.10$. Average and local baryon densities are given in units of fm$^{-3}$.}
 \label{Table5}
 \end{table}
\begin{figure}[h]
 \begin{center}
  \includegraphics[width=0.47\textwidth]{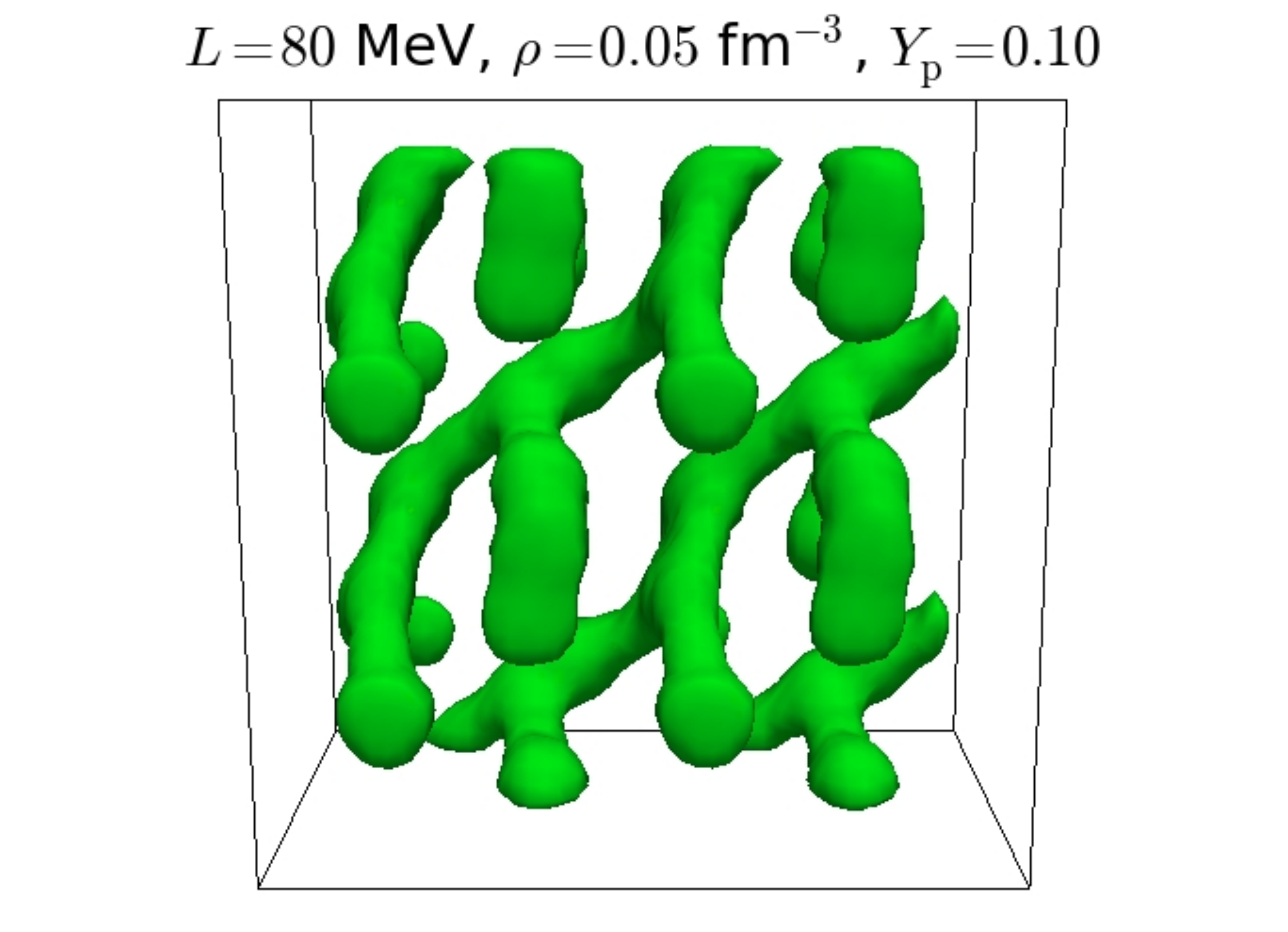}
  \caption{(Color online)
  Four periodic copies of pasta structure with $Y_{\rm p} = 0.10$, $\rho = 0.05$ fm$^{-3}$ for model
  with $L = 80$ MeV are put together for a simulation box posed at a different angle for better visualization.
  The box dimensions are $68.4 \times 68.4 \times 34.2$ fm, and
  the isosurface of proton density at $\rho_{\rm p} = 0.02$ fm$^{-3}$ is plotted.}
 \label{Fig09}
 \end{center}
\end{figure}
In proto-neutron stars, neutrinos are trapped for tens of seconds in
the hot and dense nuclear medium\,\cite{Shen:2012fq}. As neutrinos
diffuse out of the PNS, the proton fraction in this beta equilibrium
thermal matter also evolves. Therefore it is useful to explore a
large range of proton fractions in the nuclear pasta formation.
Notice that we use zero temperature in all of our simulations,
whereas in reality, the temperature in the supernova environment can
be from a few MeV to as high as $k_{\rm B} T = 10$ MeV and even
more. 

At low densities, both models again feature similar geometries (see
Fig.\,\ref{Fig08}). We observe 8 elongated nuclei randomly located
within the simulation box. The size of these structures vary from
one another, and the average charge of an individual structure is
$\langle Z \rangle \approx 25$. These structures significantly
differ from the unstable neutron-rich nuclear isotopes with the same
proton number. For example, the most neutron-rich terrestrial
radioactive Mn isotope (with $Z = 25$) known today has $N = 44$
neutrons. Surprisingly, the Coulomb frustration at $\rho = 0.03$
fm$^{-3}$ enables the formation of elongated nuclei with an average
neutron number of $N \approx 145$ ($L = 40$ MeV) or $176$ ($L = 80$
MeV). The concentration of \textit{free} neutrons at this density
now strongly depends on the interaction model (see Table
\ref{Table5}). Although a significant fraction of neutrons carry
positive kinetic energies, the overall energy of the ground state in
this system remains negative.

At $\rho = 0.04$ fm$^{-3}$, both models exhibit a very similar
geometry: two nuclei within the simulation box fuse together to form
one long rod-like structure. Thus a total of four superdeformed
rod-like nuclei are formed. As the density increases, at $\rho =
0.05$ fm$^{-3}$, we observe that rod-like structures arrange
themselves in a net-like structure for the model with $L = 40$ MeV.
Notice such structure was also observed for $Y_{\rm p} = 0.05$ but
at $\rho = 0.07$ fm$^{-3}$. On the other hand, for the system with
$L = 80$ MeV we observe a structure that resembles \textit{fibrous
roots}. To have a better view of this structure, in particular, we
show four periodic copies of the isosurface of proton densities
along two directions, mainly $x$ and $y$, using the fact that our
simulation volume is periodic. The resulting isosurface of proton
densities are plotted in Fig.\,\ref{Fig09}. Since the existence of
many low-energy configurations is the benchmark of frustrated
systems, we believe that this structure in particular could be in a
metastable state. We expect that the true ground state is a
$Y$-shaped junction that forms the backbone of a branched network of
many frustrated systems such as the low-dimensional magnetic
systems. Next, at $\rho = 0.06$ fm$^{-3}$ we observe almost
identical net-like structures in both models. At an even higher
densities the \textit{threads} of these nets structures get thicker
as a result of compression and they turn into the complex shapes
previously referred to as rod(3) structures\,\cite{Newton:2009zz,
Sebille:2009zz, Pais:2012js, Schuetrumpf:2012cj}. Notice this
structure continue to exist in the model with soft symmetry energy
even at $\rho = 0.09$ fm$^{-3}$, whereas the pasta structure almost
disappears for $L = 80$ MeV. And finally, at $\rho = 0.10$ fm$^{-3}$
both systems assume uniform phase.

\subsubsection{Systems with $Y_{\rm p} = 0.20$}
\begin{figure*}[h]
 \begin{center}
  \includegraphics[width=0.47\textwidth]{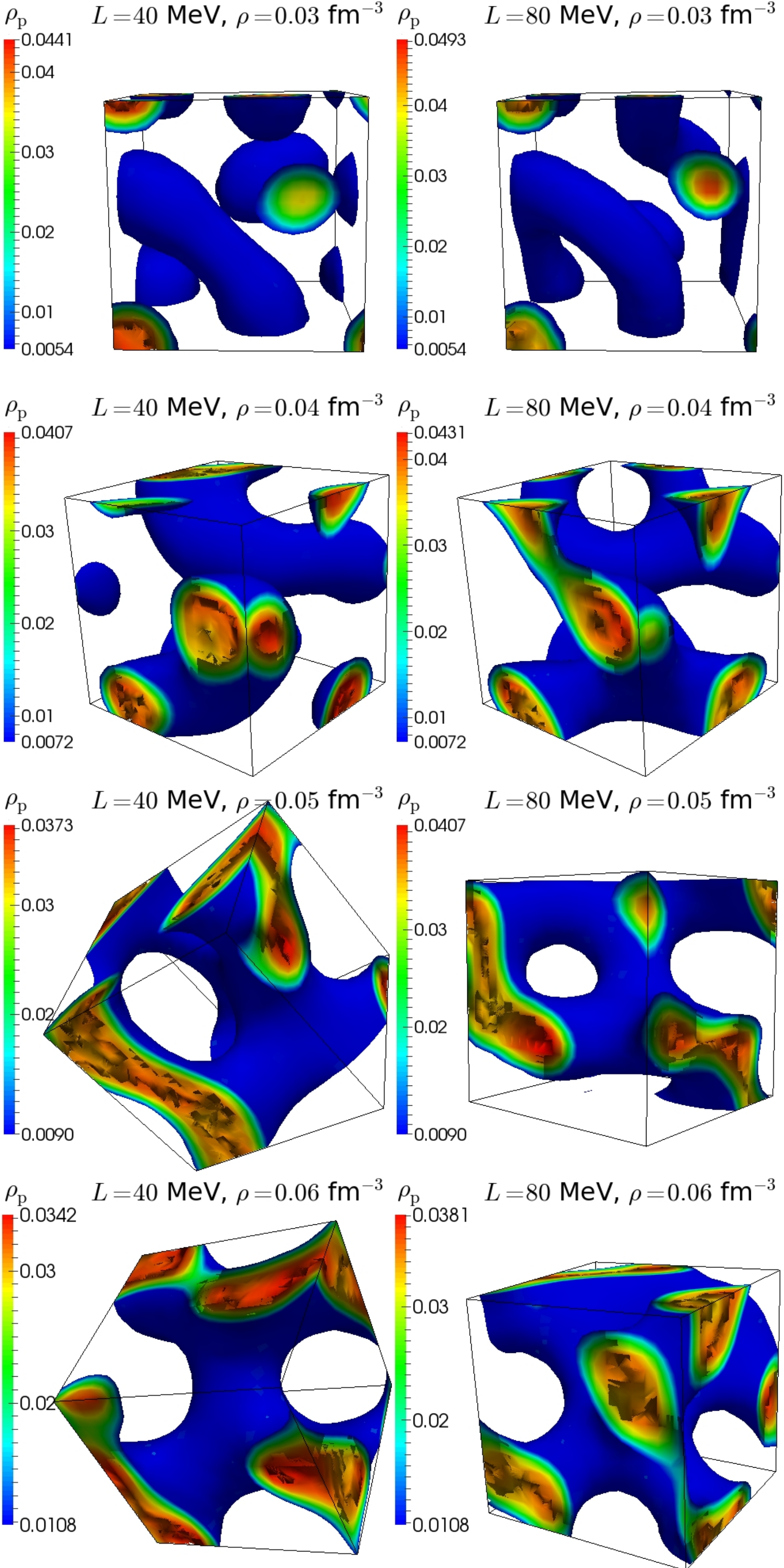}
        \hspace{0.1cm}
  \includegraphics[width=0.47\textwidth]{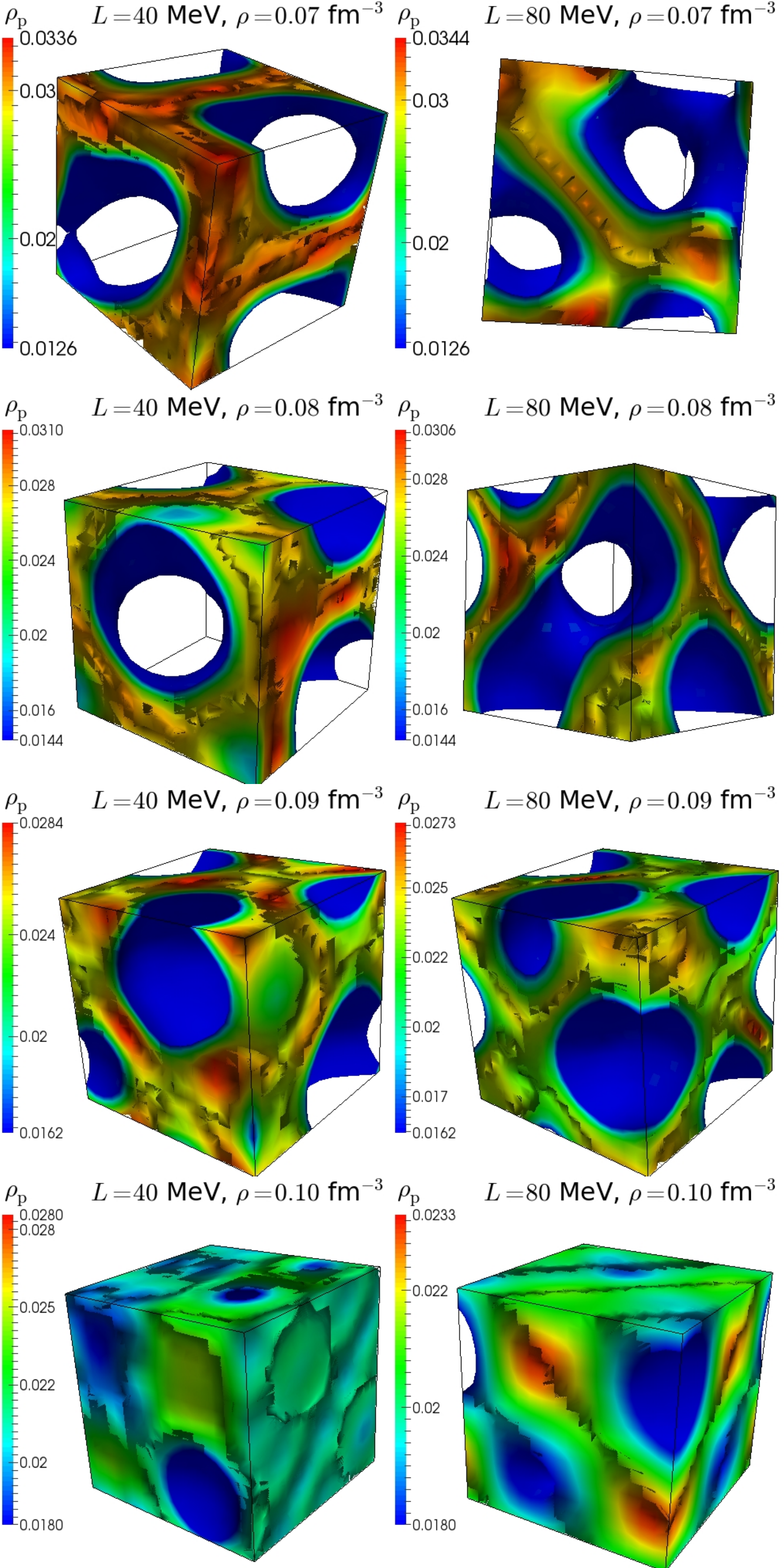}
  \caption{(Color online)
  Isosurface of proton densities are plotted using the same presciription as
  in Fig.\,\ref{Fig05} except now the proton fraction of the system is $Y_{\rm p} = 0.20$.}
 \label{Fig10}
 \end{center}
\end{figure*}
  \begin{table}[b]
  \begin{tabular}{|c|l|c|c|c|c|c|}
    \hline
    $\rho$  & Model                     & $E_{\rm tot}$ (MeV) & $\rho_{\rm tot}^{\rm min}$ & $\rho_{\rm tot}^{\rm max}$ & $N_{\rm f}$ & $Y_{\rm p}^{\star}$ (\%)  \\
    \hline
    \hline
    $0.03$  & \mbox{\tt UNEDF1}             & $-5.150$            & $0.0046$                   & $0.1438$                   & $542$       & $24.13$ \\
            & \mbox{\tt UNEDF1}$^{\star}$   & $-5.394$            & $0.0036$                   & $0.1426$                   & $394$       & $22.15$ \\
    \hline
    $0.04$  & \mbox{\tt UNEDF1}             & $-5.247$            & $0.0065$                   & $0.1382$                   & $528$       & $23.92$ \\
            & \mbox{\tt UNEDF1}$^{\star}$   & $-5.537$            & $0.0062$                   & $0.1349$                   & $396$       & $22.17$ \\
    \hline
    $0.05$  & \mbox{\tt UNEDF1}             & $-5.358$            & $0.0071$                   & $0.1354$                   & $514$       & $23.72$ \\
            & \mbox{\tt UNEDF1}$^{\star}$   & $-5.629$            & $0.0101$                   & $0.1306$                   & $400$       & $22.22$ \\
    \hline
    $0.06$  & \mbox{\tt UNEDF1}            & $-5.462$            & $0.0092$                   & $0.1296$                   & $504$       & $23.58$ \\
            & \mbox{\tt UNEDF1}$^{\star}$   & $-5.731$            & $0.0153$                   & $0.1274$                   & $406$       & $22.30$ \\
    \hline
    $0.07$  &\mbox{\tt UNEDF1}             & $-5.565$            & $0.0105$                   & $0.1302$                   & $512$       & $23.70$ \\
            & \mbox{\tt UNEDF1}$^{\star}$   & $-5.800$            & $0.0209$                   & $0.1213$                   & $436$       & $22.68$ \\
    \hline
    $0.08$  & \mbox{\tt UNEDF1}             & $-5.662$            & $0.0143$                   & $0.1253$                   & $502$       & $23.56$ \\
            & \mbox{\tt UNEDF1}$^{\star}$   & $-5.864$            & $0.0284$                   & $0.1161$                   & $464$       & $23.04$ \\
    \hline
    $0.09$  & \mbox{\tt UNEDF1}             & $-5.763$            & $0.0182$                   & $0.1194$                   & $502$       & $23.56$ \\
            &\mbox{\tt UNEDF1}$^{\star}$   & $-5.899$            & $0.0370$                   & $0.1107$                   & $508$       & $23.64$ \\
    \hline
    $0.10$  & \mbox{\tt UNEDF1}            & $-5.851$            & $0.0329$                   & $0.1222$                   & $460$       & $22.99$ \\
            & \mbox{\tt UNEDF1}$^{\star}$   & $-5.897$            & $0.0887$                   & $0.1077$                   & $562$       & $24.42$ \\
   \hline
  \end{tabular}
 \caption{Some bulk properties of nuclear pasta with an average proton fraction of
          $Y_{\rm p} = 0.20$. Average and local baryon densities are given in units of fm$^{-3}$.}
 \label{Table6}
 \end{table}
Let us now analyze the more widely studied case of systems with
larger proton fractions. Such systems display a rich-variety of
nuclear pasta even at high sub-saturation densities. For example,
even at $\rho = 0.10$ fm$^{-3}$ the density contrast in the system
is as large as $\Delta \rho = 0.089$ fm$^{-3}$ for the model with
soft symmetry energy (see Fig.\,\ref{Fig07} and
Table\,\ref{Table6}). Although the overall binding energy of the
system is negative, there are still some \textit{free} neutrons
found in this system with $Y_{\rm p} = 0.20$ (see
Table\,\ref{Table6}). Nevertheless, the fractional population of
free neutrons are much less than found before in systems with lower
proton fractions. The corresponding effective proton fractions
therefore do not deviate very much from 20\%. All pasta structures
are energetically very close to one another, yet we observe
structures that are radically different in topology. Indeed, it has
been first speculated by Ref.\,\cite{Hashimoto:1984} that the
transition from the highly ordered crystal to the uniform phase must
proceed through a series of changes in the dimensionality and
topology only that depends on density but not on total energy. We
also observe that the dependence on the symmetry energy is
significantly reduced both in total energies and in topology, even
though the system is still relatively very neutron-rich.

At $\rho = 0.03$ fm$^{-3}$ we no longer observe a system purely made
of nuclei (gnocchi phase). Instead we observe a coexistence of
nuclei and rod-like structures. When the model with the soft
symmetry energy is used we observe two nuclei and one rod in the
simulation volume. However, for $L=80$ MeV we observe just one
nucleus and a rod structure that is bent to assume a disconnected
\textit{hook-shaped} structure. At $\rho = 0.04$ fm$^{-3}$, the
first system now assumes a connected hook-shaped structures that
make a wave pattern, whereas the latter one assumes a structure that
resembles donuts which are connected through $Y$-junctions. The
lowest non-zero local baryon densities shown in Table\,\ref{Table6}
correspond to the density of background free neutron gas.

As we progressively increase the density, at $\rho = 0.05$ fm$^{-3}$
and $\rho = 0.06$ fm$^{-3}$, both systems proceed into having the
donut-like shapes with less spatial separations.  At higher
densities the size of the openings become smaller making a
transition to cylindrical holes at densities of $0.08$ fm$^{-3}$ and
eventually leading to spherical bubbles for models with the soft
symmetry energy. A similar phase transition between pasta states is
observed for models with the stiff symmetry energy, however the
system becomes uniform at much lower densities. As can be seen from
Table\,\ref{Table6} at the average baryon density of $\rho = 0.10$
fm$^{-3}$,  where the local deviation of the density within the
simulation box is no more than $\Delta \rho = 0.019$ fm$^{-3}$. This
result is also depicted in the lower right panel of
Fig.\,\ref{Fig10}.

\subsubsection{Systems with $Y_{\rm p} = 0.30$}
\begin{figure*}[h]
 \begin{center}
  \includegraphics[width=0.47\textwidth]{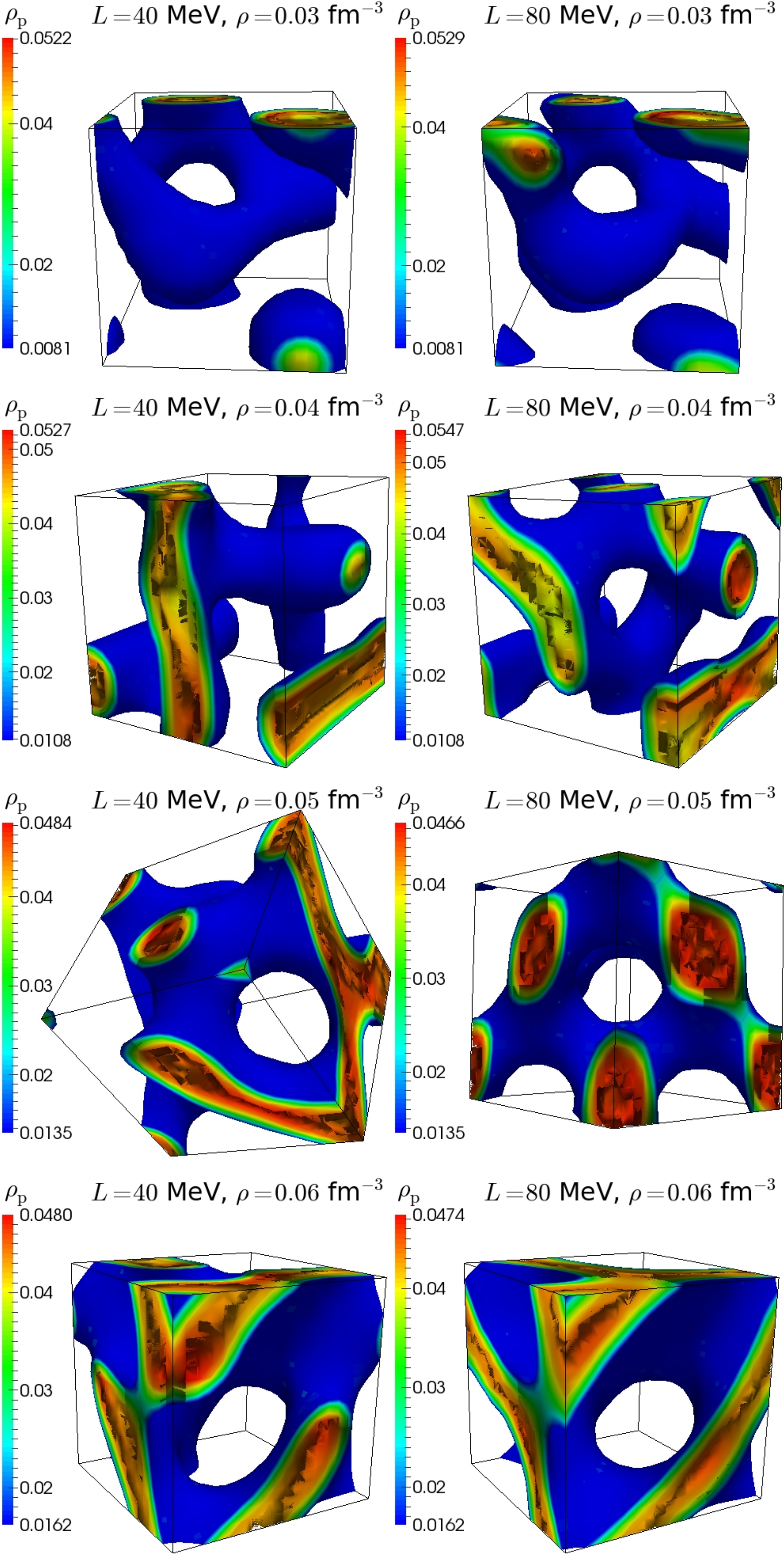}
        \hspace{0.1cm}
  \includegraphics[width=0.47\textwidth]{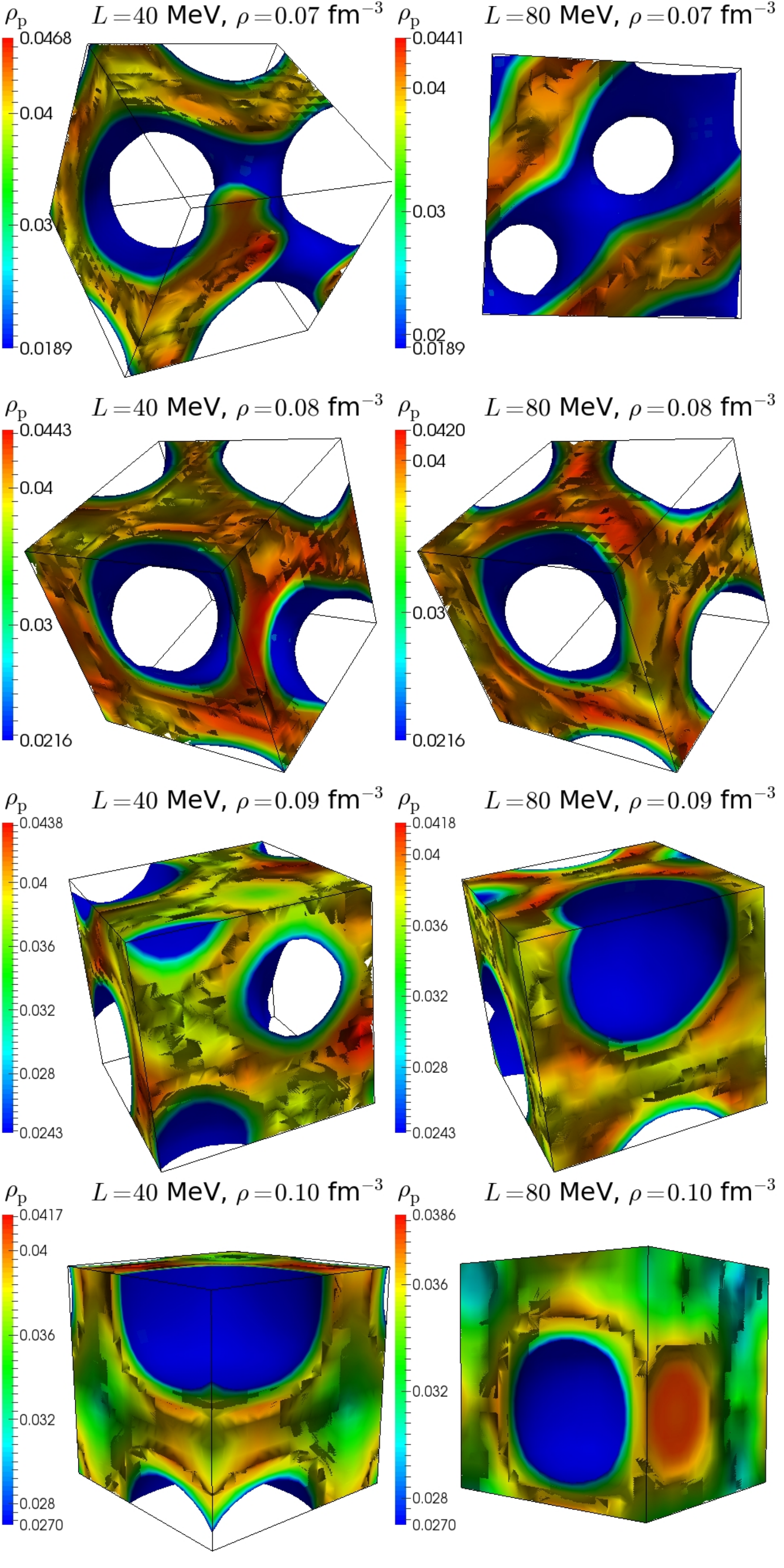}
  \caption{(Color online)
  Isosurface of proton densities are plotted using the same presciription as
  in Fig.\,\ref{Fig05} except now the proton fraction of the system is $Y_{\rm p} = 0.30$.}
 \label{Fig11}
 \end{center}
\end{figure*}
  \begin{table}[b]
  \begin{tabular}{|c|l|c|c|c|c|c|}
    \hline
    $\rho$  & Model                     & $E_{\rm tot}$ (MeV)            & $\rho_{\rm tot}^{\rm min}$ & $\rho_{\rm tot}^{\rm max}$   \\
    \hline
    \hline
    $0.03$  & \mbox{\tt UNEDF1}            & $-8.794$                       & $0.0000$                   & $0.1489$                     \\
            & \mbox{\tt UNEDF1}$^{\star}$   & $-8.782$                       & $0.0000$                   & $0.1442$                     \\
    \hline
    $0.04$  & \mbox{\tt UNEDF1}             & $-9.050$                       & $0.0000$                   & $0.1491$                     \\
            & \mbox{\tt UNEDF1}$^{\star}$   & $-9.051$                       & $0.0000$                   & $0.1470$                     \\
    \hline
    $0.05$  & \mbox{\tt UNEDF1}             & $-9.344$                       & $0.0000$                   & $0.1423$                     \\
            & \verb"UNEDF1"$^{\star}$   & $-9.337$                       & $0.0002$                   & $0.1345$                     \\
    \hline
    $0.06$  & \mbox{\tt UNEDF1}            & $-9.621$                       & $0.0000$                   & $0.1412$                     \\
            & \mbox{\tt UNEDF1}$^{\star}$   & $-9.620$                       & $0.0000$                   & $0.1360$                     \\
    \hline
    $0.07$  & \mbox{\tt UNEDF1}             & $-9.881$                       & $0.0001$                   & $0.1406$                     \\
            & \mbox{\tt UNEDF1}$^{\star}$   & $-9.843$                       & $0.0002$                   & $0.1298$                     \\
    \hline
    $0.08$  & \mbox{\tt UNEDF1}           & $-10.133$                      & $0.0002$                   & $0.1360$                     \\
            & \mbox{\tt UNEDF1}$^{\star}$   & $-10.085$                      & $0.0004$                   & $0.1267$                     \\
    \hline
    $0.09$  & \mbox{\tt UNEDF1}            & $-10.371$                      & $0.0001$                   & $0.1351$                     \\
            & \mbox{\tt UNEDF1}$^{\star}$   & $-10.308$                      & $0.0003$                   & $0.1265$                     \\
    \hline
    $0.10$  & \mbox{\tt UNEDF1}            & $-10.601$                      & $0.0001$                   & $0.1306$                     \\
            & \mbox{\tt UNEDF1}$^{\star}$   & $-10.530$                      & $0.0005$                   & $0.1194$                     \\
   \hline
  \end{tabular}
 \caption{Some bulk properties of nuclear pasta with an average proton fraction of
          $Y_{\rm p} = 0.30$. Average and local baryon densities are given in units of fm$^{-3}$.}
 \label{Table7}
 \end{table}

Turning to increasingly symmetric matter, in Fig.\,\ref{Fig11} we
display the isosruface of the proton densities of various pasta
phases for $Y_{\rm p} = 0.30$ using both models with $L = 40$ and $L
= 80$ MeV. It is observed that such systems exhibit a series of many
complex geometries. We no longer observe spherical nuclei at an
average density of 0.03 fm$^{-3}$, which was chosen as the starting
point of our simulations. Obviously, the gnocchi phase must have
formed at an even lower density for $Y_{\rm p} = 0.30$. At the
lowest density considered in our simulation we observe fibrous
root-like structures, at 0.04 fm$^{-3}$ we observe rod(3)
structures. At densities of 0.05 and 0.06 fm$^{-3}$ the pasta system
is composed of circular perforated complex systems, at 0.07 and 0.08
fm$^{-3}$ the nuclear pasta transitions to the bucatini phase, and
finally at 0.09 and 0.10 fm$^{-3}$ they form the Swiss cheese. The
pasta systems are strongly bound with binding energies ranging from
$-8.8$ MeV for systems with average baryon density of 0.03 fm$^{-3}$
to $-10.1$ MeV for systems with $\rho =0.10$ fm$^{-3}$ (see
Table\,\ref{Table7}).

All neutrons strongly participate in forming the pasta structure,
and there are no free neutrons left in the system. Thus the neutron
gas background that was making the lowest density of the simulation
box in the previous systems with lower proton fractions now simply
vanishes. The vanishing of the neutron gas background for $Y_{\rm p}
> 0.29$ was also obtained earlier Ref.\,\cite{Schuetrumpf:2015nza}.
For this and larger proton fractions one can either plot the
isosurface of proton densities or total densities that are both
visually indistinguishable.

Perhaps the most interesting aspect of this system is to notice that
the dependence on the nuclear symmetry energy has now become less
prominent. The binding energies in all configurations are very
close. The similarity of density contrasts for these configurations
as predicted by both models suggest that the pasta structures should
also be close to one another, which is confirmed by comparing them
as displayed in Fig.\,\ref{Fig11}. Thus while the symmetry energy
plays a significant role for the nuclear pasta formation in the
neutron star crust and for the the regions of Supernovae with low
proton fractions its role becomes insignificant for pasta formation
at $Y_{\rm p} \gtrsim 0.30$. This result is one of the important
findings of our work.

\subsubsection{Systems with $Y_{\rm p} = 0.40$}
\begin{figure*}[h]
 \begin{center}
  \includegraphics[width=0.47\textwidth]{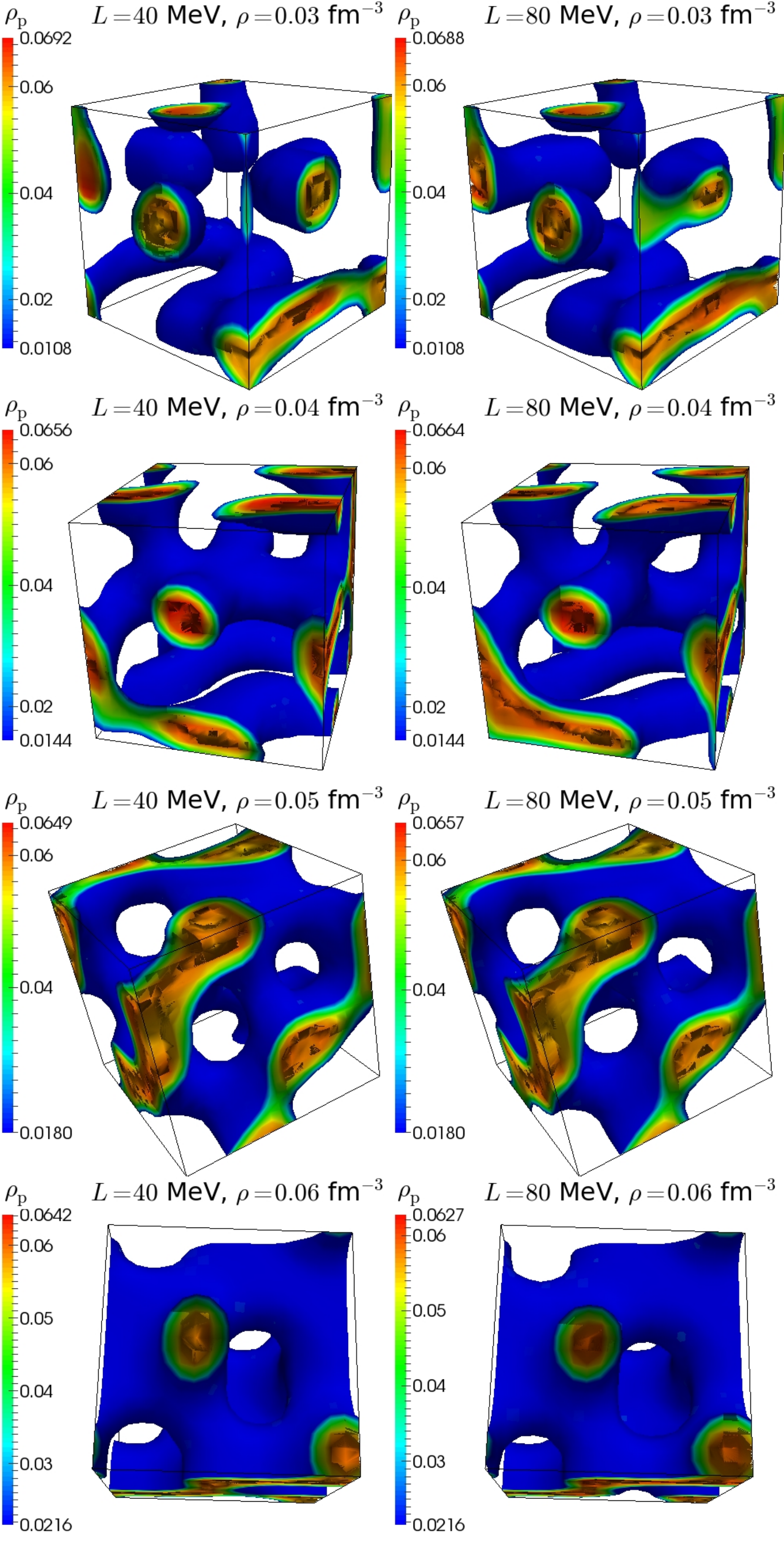}
        \hspace{0.1cm}
  \includegraphics[width=0.47\textwidth]{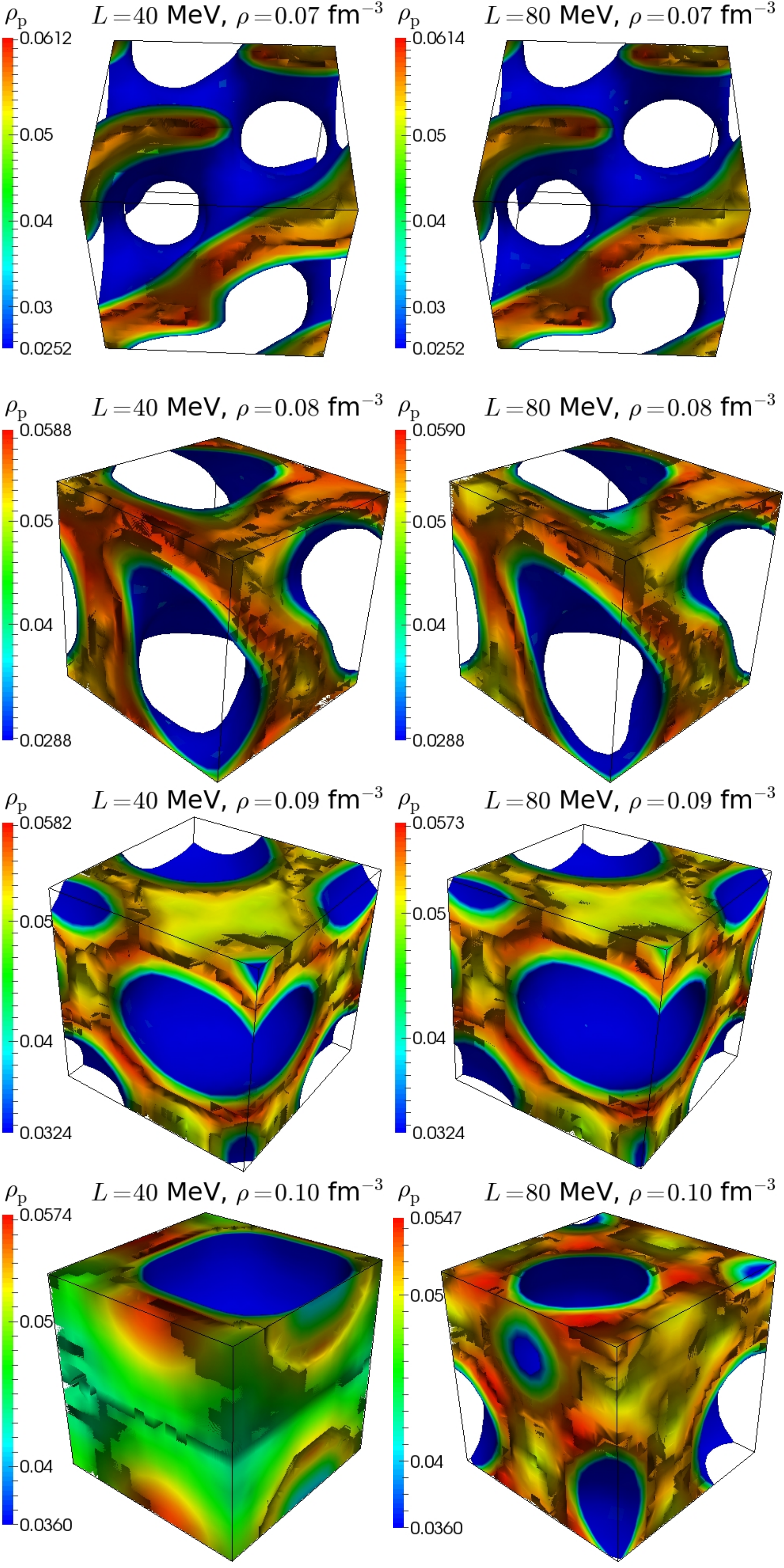}
  \caption{(Color online)
  Isosurface of proton densities are plotted using the same presciription as
  in Fig.\,\ref{Fig05} except now the proton fraction of the system is $Y_{\rm p} = 0.40$.}
 \label{Fig12}
 \end{center}
\end{figure*}
  \begin{table}[b]
  \begin{tabular}{|c|l|c|c|c|c|c|}
    \hline
    $\rho$  & Model                     & $E_{\rm tot}$ (MeV)            & $\rho_{\rm tot}^{\rm min}$ & $\rho_{\rm tot}^{\rm max}$   \\
    \hline
    \hline
    $0.03$  & \mbox{\tt UNEDF1}             & $-11.076$                       & $0.0000$                   & $0.1605$                     \\
            & \mbox{\tt UNEDF1}$^{\star}$   & $-11.050$                       & $0.0000$                   & $0.1585$                     \\
    \hline
    $0.04$  &\mbox{\tt UNEDF1}            & $-11.442$                       & $0.0000$                   & $0.1559$                     \\
            &\mbox{\tt UNEDF1}$^{\star}$   & $-11.394$                       & $0.0000$                   & $0.1549$                     \\
    \hline
    $0.05$  & \mbox{\tt UNEDF1}             & $-11.780$                       & $0.0000$                   & $0.1532$                     \\
            & \mbox{\tt UNEDF1}$^{\star}$   & $-11.756$                       & $0.0000$                   & $0.1531$                     \\
    \hline
    $0.06$  & \mbox{\tt UNEDF1}            & $-12.173$                       & $0.0000$                   & $0.1529$                     \\
            & \mbox{\tt UNEDF1}$^{\star}$   & $-12.145$                       & $0.0000$                   & $0.1486$                     \\
    \hline
    $0.07$  & \mbox{\tt UNEDF1}           & $-12.516$                       & $0.0000$                   & $0.1472$                     \\
            & \mbox{\tt UNEDF1}$^{\star}$   & $-12.482$                       & $0.0000$                   & $0.1457$                     \\
    \hline
    $0.08$  & \mbox{\tt UNEDF1}             & $-12.873$                      & $0.0000$                   & $0.1426$                     \\
            & \mbox{\tt UNEDF1}$^{\star}$   & $-12.830$                      & $0.0000$                   & $0.1412$                     \\
    \hline
    $0.09$  &\mbox{\tt UNEDF1}          & $-13.194$                      & $0.0000$                   & $0.1395$                     \\
            & \mbox{\tt UNEDF1}$^{\star}$   & $-13.156$                      & $0.0000$                   & $0.1377$                     \\
    \hline
    $0.10$  & \mbox{\tt UNEDF1}            & $-13.501$                      & $0.0000$                   & $0.1363$                     \\
            & \mbox{\tt UNEDF1}$^{\star}$   & $-13.504$                      & $0.0000$                   & $0.1334$                     \\
   \hline
  \end{tabular}
 \caption{Some bulk properties of nuclear pasta with an average proton fraction of
          $Y_{\rm p} = 0.40$. Average and local baryon densities are given in units of fm$^{-3}$.}
 \label{Table8}
 \end{table}

Finally, we study the case of $Y_{\rm p} = 0.40$. This proton
fraction is roughly comparable to that found in the collapsing dense
core of a supernovae, before the matter gets heated further by the
shock wave. As confirmed in the previous subsection and given the
fact that the matter is close to being isospin symmetric, the role
of the symmetry energy becomes negligible. The maximum local density
in the system is equal to that of the nuclear saturation density,
$\sim 0.16$ fm$^{-3}$. At densities of 0.03 and 0.04 fm$^{-3}$ the
pasta system is made of connected rod structures (see
Fig.\,\ref{Fig12}). At $\rho = 0.05$ fm$^{-3}$, rods merge to form
complex structures with circular openings and at $\rho = 0.06$
fm$^{-3}$ they form states that closely resemble the perforated
parallel plates, which are now connected along their normal
direction. Again, as density increases, the matter forms cylindrical
holes at $0.07$ and $0.08$ fm$^{-3}$, and finally spherical bubbles
are observed at higher densities.

The nuclear pasta at $Y_{\rm p} = 0.40$ is strongly bound, no
neutron background exists as in the case of $Y_{\rm p} = 0.30$. The
binding energy per nucleon is much smaller than that of the uniform
matter. For example, at $\rho = 0.03$ fm$^{-3}$ we have $E/A =
-11.05$ MeV, whereas the uniform nuclear matter predicts an almost
twice smaller value of $E/A = -5.38$ MeV. When symmetric
nuclear matter (SNM) is considered a similarly large difference in
the binding energies per nucleon would obviously be expected between
the uniform nuclear matter and the nuclear pasta. Given this fact, a
word of caution on the definition of the symmetry energy or the SNM
is in order. The symmetry energy $S(\rho)$ which is defined as the
coefficient of expansion of the binding energy per nucleon,
$\epsilon \equiv E/A$,
\begin{equation}
\epsilon(\rho, \alpha) = \epsilon(\rho, 0) + S(\rho) \alpha^2 +
\ldots \ ,
\end{equation}
where $\alpha= (\rho_{\rm n}-\rho_{\rm p})/(\rho_{\rm n}+\rho_{\rm
p})$ is the isospin asymmetry, usually represents the energy cost
per nucleon of changing all the protons in SNM into neutrons. We
should be wary about the latter description because the ground state
of matter at sub-saturation densities is not that of a uniform
matter. Therefore in this description where the usage of SNM
appears, one should explicitly state that a hypothetical
\textit{uniform} nuclear matter has been considered, which can
easily cluster if left alone.

\subsection{Sensitivity of the ``Ground State" to the Initial Configurations}
\begin{figure}[h]
  \includegraphics[width=0.45\textwidth]{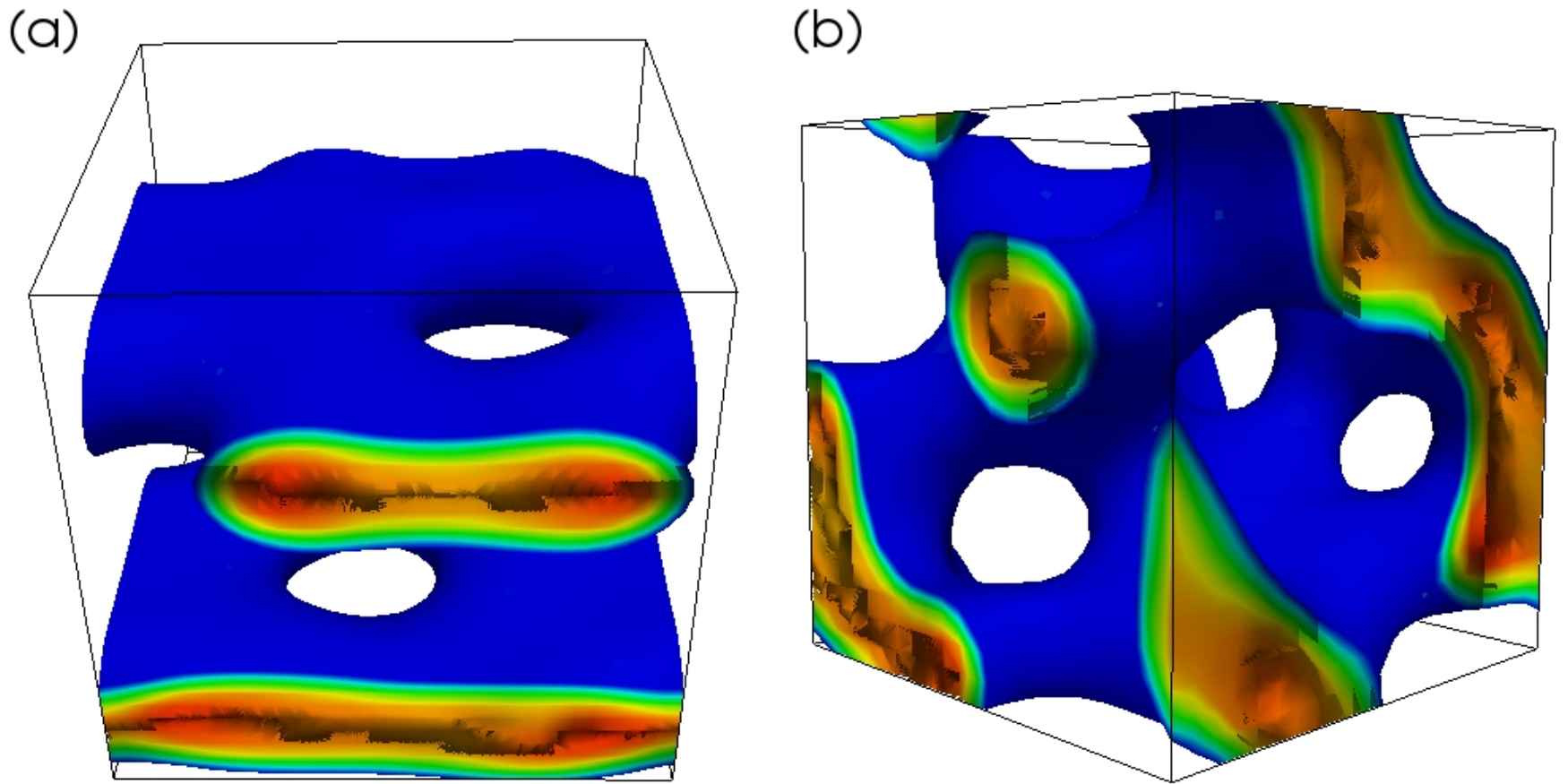}
  \caption{(Color online)
  Nuclear pasta phases at $\rho = 0.05$ fm$^{-3}$ and $Y_{\rm p} = 0.40$ that
  started out from two different initial configurations with nucleons randomly distributed in the box
  and the grid spacing of (a) $\Delta x = 1.00$ fm and (b) $\Delta x = 1.42$ fm.}
 \label{Fig13}
\end{figure}
\begin{figure}[h]
  \includegraphics[width=0.50\textwidth]{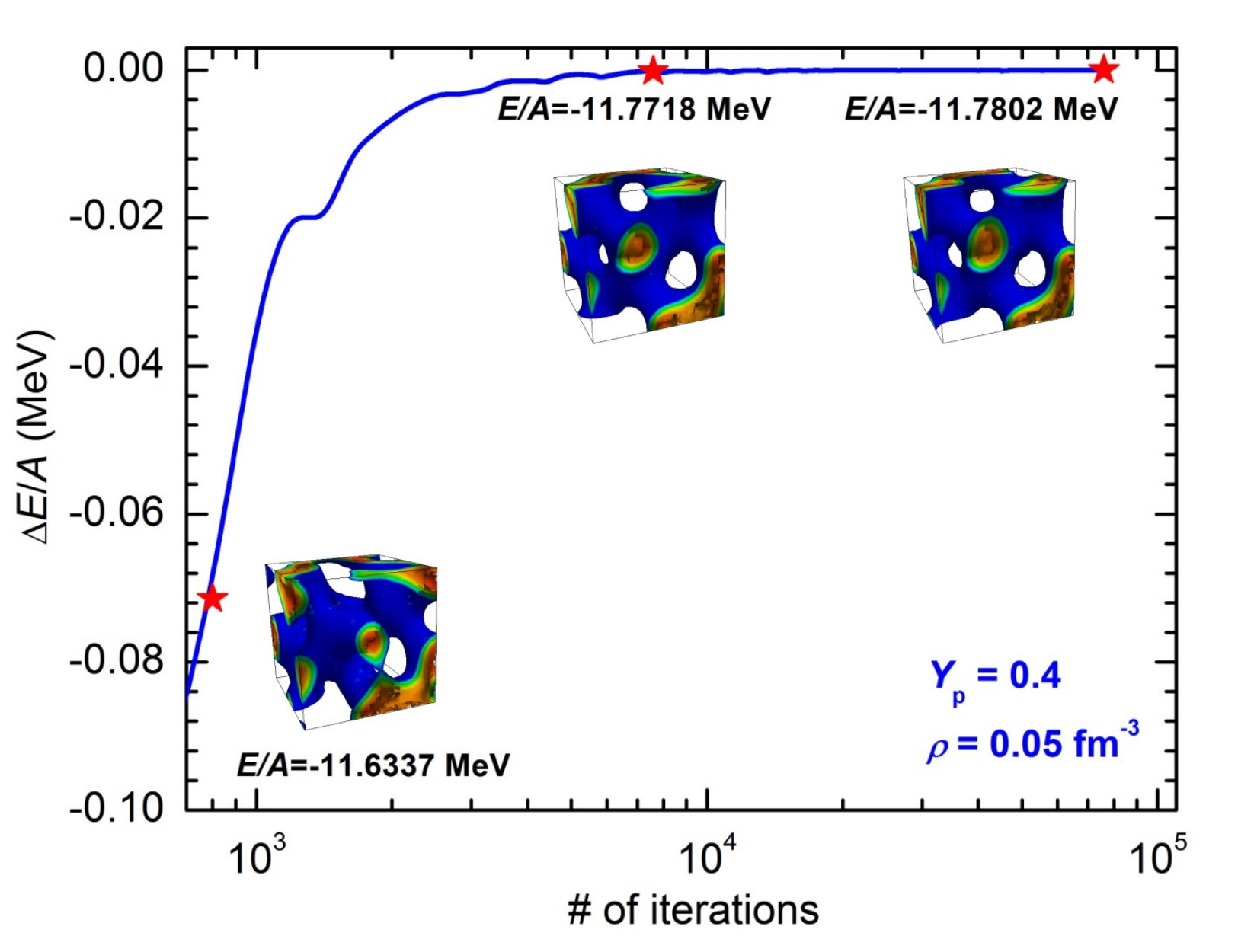}
  \caption{(Color online)
  Energy difference per 200 iterations versus the number of iterations is plotted for a total of 76,000 iterations. The inset pasta phases correspond
  the simulation phases at 800, 7600, and 76000 iterations, respectively.}
 \label{Fig14}
\end{figure}

\begin{figure}[p]
 \begin{center}
  \includegraphics[width=0.4\textwidth]{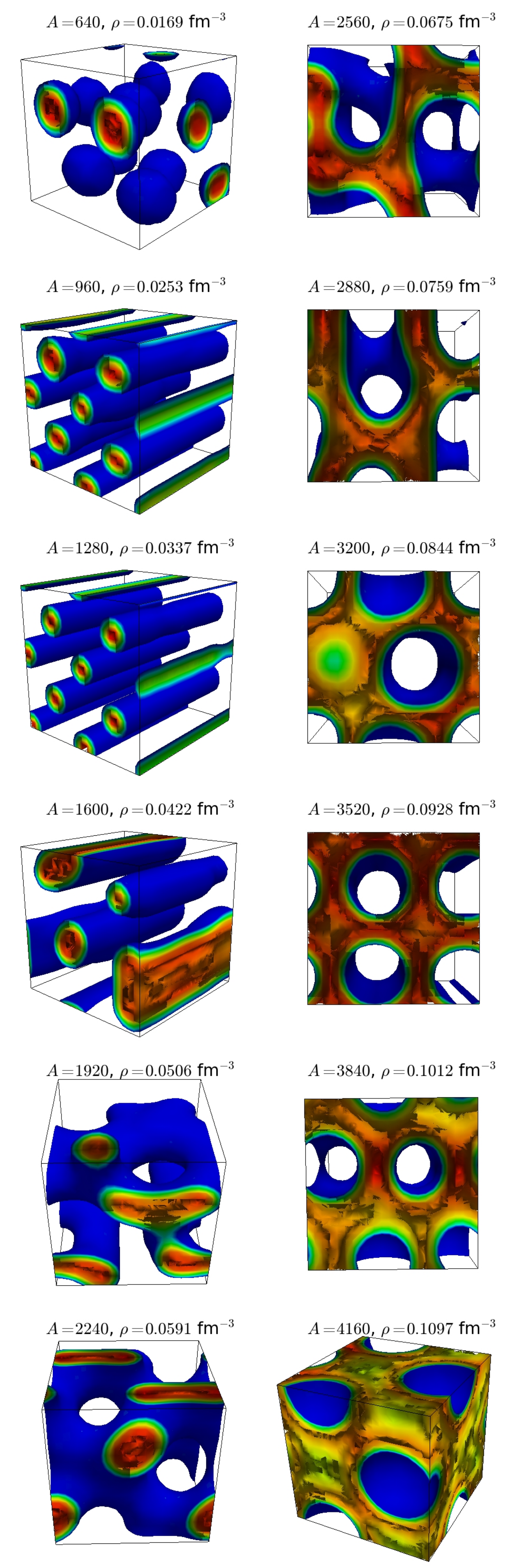}
 \end{center}
   \caption{(Color online)
  Nuclear pasta simulations with proton fractions of $Y_{\rm p} = 0.40$ for a cubic volume of fixed side $a=33.6$ fm containing
  $640 < A < 4180 $ nucleons. All systems are initialized
  with nucleons distributed randomly to form eight identical rods aligned on a face-centered site.}
 \label{Fig15}
\end{figure}

\begin{figure}[p]
 \begin{center}
  \includegraphics[width=0.45\textwidth]{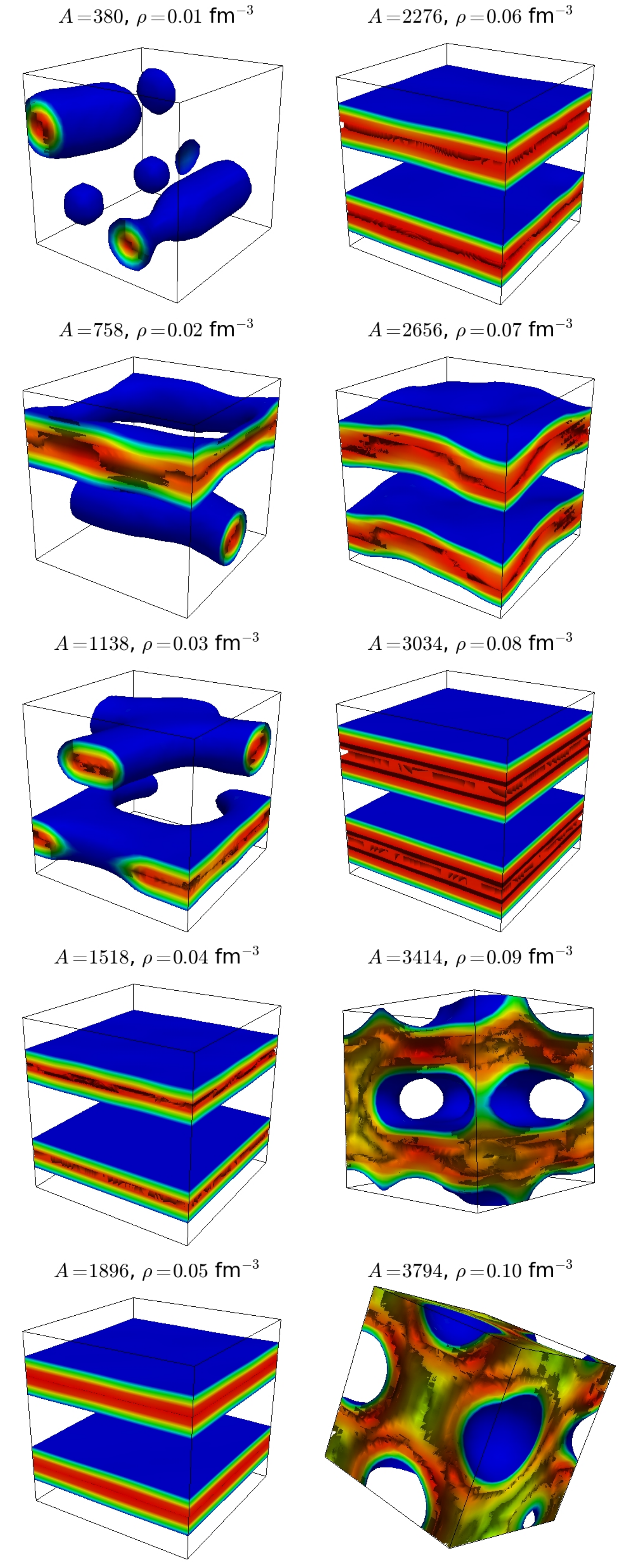}
 \end{center}
   \caption{(Color online)
  Nuclear pasta simulations with proton fractions of $Y_{\rm p} = 0.40$ at average baryon densities of $0.01 < \rho < 0.10$
  fm$^{-3}$ corresponding to a cubic volume with $a \approx 33.6$ fm. All systems are initialized
  with nucleons distributed randomly to form two identical parallel plates.}
 \label{Fig16}
\end{figure}

\begin{figure}[p]
 \begin{center}
\includegraphics[width=1.0\columnwidth,angle=0]{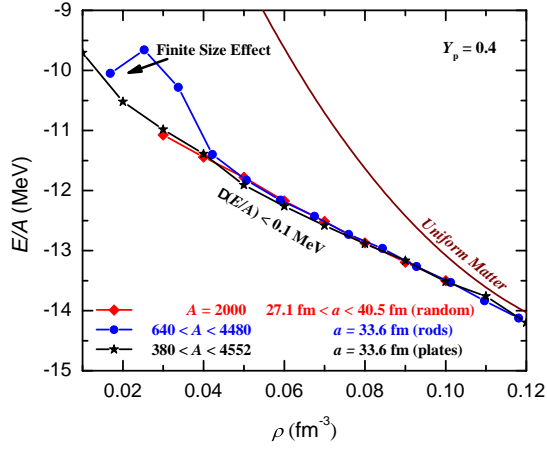}
 \end{center}
   \caption{(Color online)
  Total energy per nucleon as a function of density for pasta structures that are obtained
  from three different initial configurations.}
 \label{Fig17}
\end{figure}
\begin{figure}[p]
 \begin{center}
\includegraphics[width=1.0\columnwidth,angle=0]{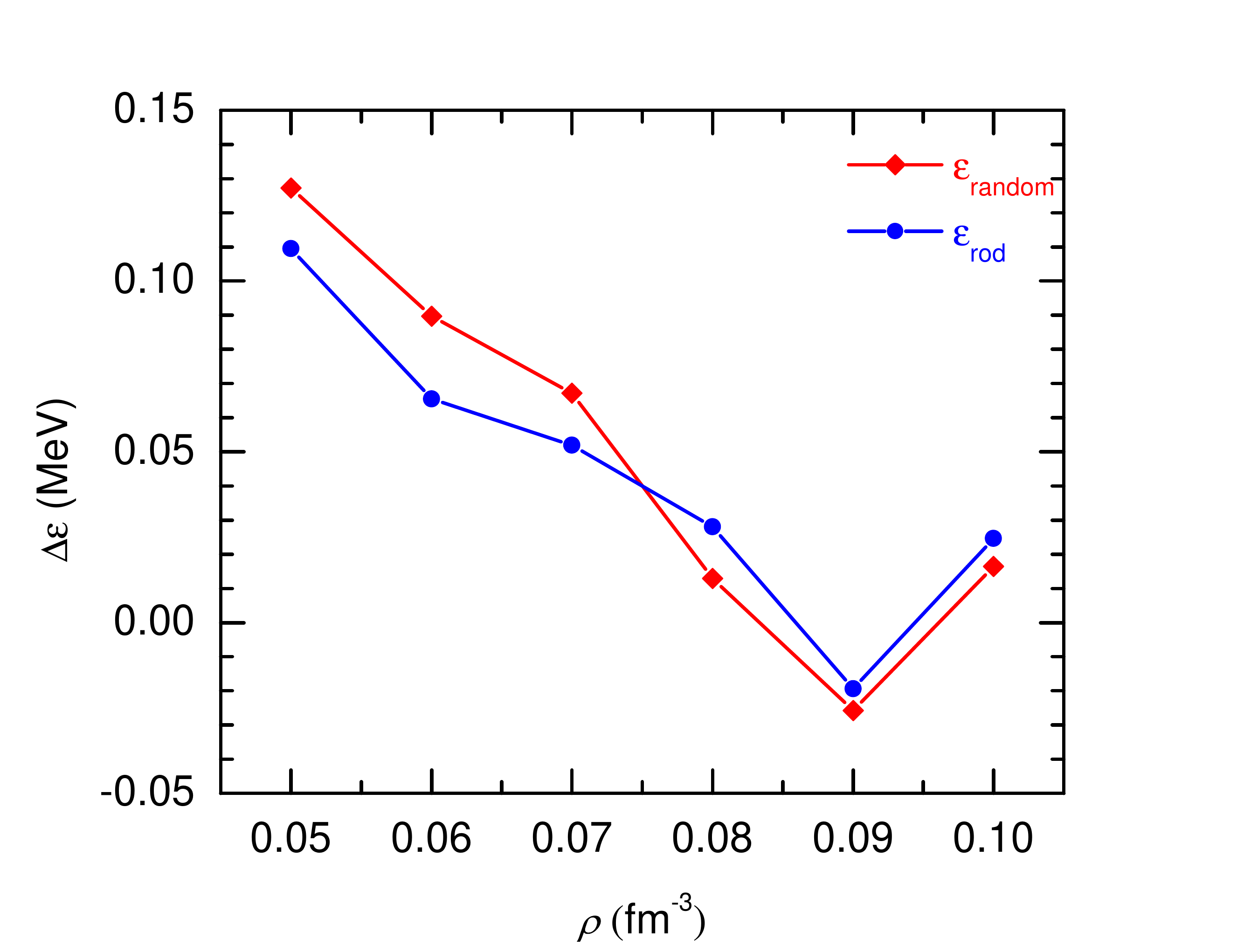}
 \end{center}
   \caption{(Color online)
  The energy differences of final configurations shown in Fig.\,\ref{Fig17}. Here
$\Delta \epsilon_{\rm random} \equiv (E/A)_{\rm random} - (E/A)_{\rm
plate}$ and  $\Delta \epsilon_{\rm rod} \equiv  (E/A)_{\rm rod} -
(E/A)_{\rm plate}$.}
 \label{Fig18}
\end{figure}
Notice that none of our simulations have produced parallel plates.
One reason is because parallel plates might have formed in a very
small density range not considered in our simulations. Indeed, using
an almost $10$ times smaller number of particles, but exploring a
density range of $0.02 < \rho <0.12$ fm$^{-3}$ with smaller steps of
$0.025$ fm$^{-3}$, Ref.\,\cite{Schutrumpf:2014vqa} has observed
parallel plates to appear within a very short density range. The
other reason is because our simulation could significantly depend on
the initial configurations of the system. In most of other previous
full quantum mechanical studies the existence of stable plate
configuration was usually confirmed by assuming that the initial
state of the system is already in the plate configuration and by
using certain guiding potentials that lead to this form. Since we
have started from a completely random distributions of nucleons, it
is not guaranteed that our final configurations are in the true
ground state of the nuclear pasta, but the solutions are driven to a
meta-stable state.

We start the analysis by comparing two identical configurations with
$\rho=0.05$ fm$^{-3}$, $Y_{\rm p} = 0.40$, and $A =2000$ that have
started from different random initial configurations and different
grid spacings. At the final converged stated we obtained $E/A =
-11.852$ MeV and $E/A = -11.780$ MeV, respectively. While these
states have similar energies, the final pasta shapes are not quite
identical. The first one gives two parallel plates with
wholes---nuclear waffle---whereas the second one gives perforated
plates with complex 3D connections, see Fig.\,\ref{Fig13}. There
could be two reasons behind this difference. First reason is that
the grid spacings in two simulations were different, with the first
one being a fine grid spacing of $\Delta x = 1.00$ fm, whereas with
the second one being $\Delta x = 1.42$ fm. Our energy difference of
$0.072$ MeV at first suggests that perhaps a finer grid spacing
should be sought in the future simulations. However, earlier in
Section \ref{NucIntNSE} and Fig.\,\ref{Fig02} we have shown that the
dependence on the grid spacing should be minimal with energy
difference of less than $0.007$ MeV if started from the same initial
configuration. Whereas the difference of $0.072$ MeV is still tiny
(about $0.6\%$ only), the observed pasta topologies are quite
different. The second reason for this could therefore be that the
final state of the system is very sensitive to the initial
configurations. In Fig.\,\ref{Fig14} we compare intermediate pasta
states during the convergence at various iteration points. For the
first $\sim 5\,000$ iterations the simulation converges quickly, and
in the remaining $\sim 70\,000$ iterations we do not see a
significant change in both the energy and the topology of the
system. This suggests that the simulation gets trapped in a
meta-stable state after the first few thousands iterations. For
example, the energy difference of only $\Delta \epsilon_{\rm tot} =
\epsilon_{\rm tot}^{(76\,000)} - \epsilon_{\rm tot}^{(7\,600)} <
-{0.0084}$ MeV is observed in the last 68,400 iterations
corresponding to 28,700 CPU hours in the simulation runtime. This
suggests that it is not important to run the \verb"Sky3D"
simulations over about, $10\,000$ iterations, which saves a
considerable amount of CPU hours. The question then arises on how to
find the true ground state of the nuclear pasta.

To further study this in more details we have explored three
possibilities. In addition to an already discussed case with the
initial configuration of randomly distributed nucleons in the
simulation volume, we have considered two other cases with initial
configurations of: (a) parallel rods on a face-centered site (spaghetti
phase) and (b) parallel plates (lasagna phase). The MD simulations
for large proton fractions suggest that the spaghetti phase should
appear at densities of $0.02 \lesssim \rho \lesssim 0.04$ fm$^{-3}$,
whereas the lasagna phase should appear at densities of $0.05
\lesssim \rho \lesssim 0.07$ fm$^{-3}$\,\cite{Schneider:2013dwa}.
Starting out from pre-assumed spaghetti and lasagna phases we
therefore expect these pasta phases to remain stable at these
densities.

The spaghetti case is prepared as follows. We fixed the simulation
volume to be cubic with sides of $a = 33.6$ fm. The grid spacing was
fixed at $\Delta x = 1.40$ fm. A total of eight identical parallel
rods whose axes align along the z-direction and are packed in a
face-centered site were formed by randomly distributing neutrons and
protons within the rod structure. Since each rod structure contains
the same number of neutrons or protons, the total proton number $Z$
and neutron number $N$ were therefore chosen as multiples of eight.
Furthermore, since the volume of the system was fixed, the average
baryon density cannot be set arbitrarily but is determined by the
number of nucleons, $A$. We considered a total of $13$
configurations with the number of nucleons ranging from $640 \leq A
\leq 4480$. The corresponding average baryon densities are $0.0169 <
\rho < 0.1181$ fm$^{-3}$. In Fig.\,\ref{Fig15} we display our
results for these simulations. At very low densities the system
arranges itself into eight $^{32}$Ge isotopes. Notice that indeed
the spaghetti phase remains stable in agreement with the MD
simulations even when full quantum mechanical effects are
considered. Whereas there is a qualitative agreement with the
results displayed in Fig.\,\ref{Fig12} the overall topology is quite
different in two cases. The complex perforated plates with normal
connections are observed at densities of $\approx 0.05 \sim 0.06$
fm$^{-3}$. At higher densities the nuclear pasta transitions into
the bucatini phase (anti-spaghetti). It is very interesting to note
that the cylindrical holes continue to exist even at a very high
density of $\rho = 0.10$ fm$^{-3}$, where we observe a phase
coexistence between the cylindrical holes and spherical bubbles.
Recall that when we initialized the system with a randomly
distributed nucleons this phase got diminished already at $\rho =
0.09$ fm$^{-3}$ (compare with Fig.\,\ref{Fig12}). Moreover, to the
best of our knowledge, cylindrical holes at densities as large as
$0.10$ fm$^{-3}$ have never been observed in the previous
simulations that uses simplified interactions. And finally, regular
spherical bubbles (anti-gnocchi) are formed at a very high density
of $0.11$ fm$^{-3}$. The system assumes a uniform phase at higher
densities and we did not display our result here.

In the next case, we prepared our initial configurations assuming
that all nucleons are evenly distributed to make two parallel
plates. By folding Gaussians over each nucleon we constructed the
initial single-particle wave functions and solved Hartree-Fock
equations iteratively. When the simulation is converged we observe
completely different topologies than the ones observed before (see
Fig.\,\ref{Fig16}). In particular, at a very low density of $0.01$
fm$^{-3}$ we observe two super-elongated nuclei and two spherical
nuclei (that resembles baseball bat and ball). At $0.02$ fm$^{-3}$,
a phase coexistence between two types of rods is observed.
Particularly interesting is the nuclear waffle state that forms much
earlier than observed before at the density of $0.03$ fm$^{-3}$. And
the initial lasagna phase remains stable over a large density region
of $0.04 \lesssim \rho \lesssim 0.08$ fm$^{-3}$. The anti-spaghetti
phase is not observed at all within the density steps we considered
in our simulations. After developing through spherical bubbles at
0.10 fm$^{-3}$, the pasta structure completely disappears at $\rho
\gtrsim 0.11$ fm$^{-3}$.

These results obtained above are the consequence of generic features
of matter frustration that allows many different local energy
minima, hence pasta topologies. Thus we have obtained a series of
pasta geometries where matter got trapped in a quasi-ground state.
In order to determine which of these states represent the true
ground state, in Fig.\,\ref{Fig17} we plot the (quasi-) ground state
energies per nucleon as a function of average baryon density for all
three cases considered above. As evident from the figure,
energetically these pasta structures are very close to one another.
A careful observation of energies suggests that at densities of
$0.05 < \rho < 0.07$ fm$^{-3}$, for example, the system favors the
lasagna phase (See Fig.\,\ref{Fig18}). However, considering that we
explored only few possibilities, it is difficult to predict the true
ground state of the system---hence the formation of other pasta
geometries---just by comparing these energies alone.

As a final note, we would like to point out that one way to get a
time-efficient convergence is to start solving the Hartree-Fock
equations by initializing the single-particle wave functions from an
already converged classical or quantum MD simulations that have
shown to give a full qualitative picture of nuclear pasta
topologies. This will significantly reduce the simulation runtime,
which in turn allows to explore much larger simulation volumes. Our
preliminary calculations show that the ground state energies are
slightly lower when the simulation is initialized from a converged
state of classical MD simulations. Clearly, much work remains to be
done in these fronts to determine the true ground state of the
nuclear pasta.
\section{Conclusions}
\label{Conclusions}

In this work we performed large volume simulations of the nuclear
pasta using the Skyrme Hartree Fock calculations with \verb"Sky3D".
We considered a range of proton fractions with $Y_{\rm p} =0.05$,
$0.10$, $0.20$, $0.30$ and $0.40$ as well as the range of baryon
densities from $0.03 < \rho < 0.10$ fm$^{-3}$, applicable to the
nuclear matter found in the neutron star crust and supernovae. We
discussed the role of the nuclear symmetry energy in the pasta
formation and have found that it strongly impacts the nuclear pasta
geometries in the neutron star crust but has negligible effect on
the nuclear pasta in the Supernovae, where the proton fraction is
large. In particular, the crust of the neutron star contains a
larger density regions with pasta if the nuclear symmetry energy is
soft. Various nuclear pasta geometries exist even if the density
slope of the nuclear symmetry energy is as large as $L = 80$ MeV in
agreement with previous calculations. All pasta regions are found to
be filled with the neutron gas background for proton fractions
$Y_{\rm p} < 0.30$ fm. At higher proton fractions, neutron gas
background vanishes, and all neutrons in the system strongly
participate in forming the pasta topology.

Particularly interesting is the nuclear waffle formation.
Independently from the classical MD simulations, we confirmed that
the nuclear waffle state forms naturally even when full quantum
mechanical effects are considered. The existence of disconnected rod
structures with $Y$-shaped junctions hints that many of these pasta
geometries can be in the quasi-ground state. We have explored three
possible scenario in which the initial state of the system was
prepared by assuming that nucleons are randomly distributed within
(1) the full simulation volume, (2) eight parallel rods on a
face-centered site, and (3) two parallel plates. The resulting
energies of the system are very close to one another with $\Delta
E/A < 0.1$ MeV. Nevertheless, the final ``ground" state in each
scenario have distinct pasta geometries. Determination of the true
ground state requires the exploration of all possible probes which
is quite tedious in practice. As a possible alternative, we suggest
to initiate simulations from the final state of various classical or
quantum MD solutions. This way the system will be converged
time-efficiently, MD pasta geometries can be tested for stability
against density fluctuations by using full quantum simulation, and
quantitatively accurate results can be presented.

\begin{acknowledgments}
\vspace{-0.4cm} FJF and CJH are supported by the U.S. Department of
Energy (DOE) grants DE-FG02-87ER40365 (Indiana University),
DE-SC0008808 (NUCLEI SciDAC Collaboration) and by the National
Science Foundation through XSEDE resources provided by the National
Institute for Computational Sciences under grant TG-AST100014. BS is
supported by DOE grant DE-SC0008511 (NUCLEI SciDAC-3 collaboration).
This work benefited in parts from discussions at the Frontiers in
Nuclear Astrophysics meeting supported by the National Science
Foundation under Grant No. PHY-1430152 (JINA Center for the
Evolution of the Elements). The authors would like to thank the
developers of the code \verb"Sky3D" and are grateful to Indiana
University for accessing to the resources at the Big Red II
supercomputer.
\end{acknowledgments}
\vfill

\bibliography{ReferencesFJF}
\end{document}